\documentclass[letterpaper, 12pt]{article}
\usepackage{latexsym}
\usepackage{amsmath}
\usepackage{amssymb}
\usepackage[latin1]{inputenc}
\usepackage{graphicx}
\usepackage{color}
\newcommand{\U}{\textrm{U}}

\def\CN{{\cal N}}

\newcommand{\ads}[1]{{\rm AdS}_{#1}}

\newcommand{\be}{\begin{equation}}
\newcommand{\ee}{\end{equation}}
\newcommand{\bea}{\begin{eqnarray}}
\newcommand{\eea}{\end{eqnarray}}

\newcommand{\setall}{\setcounter{equation}{0}
        \setcounter{theorem}{0}}

\def\IC{\mathbb{C}}

\def\IR{\mathbb{R}}
\def\cN{{\cal N}}
\def\eM{{\mathcal M}}
\def\eH{{\mathcal H}}

\newcommand{\labell}[1]{\label{#1}\qquad_{#1}} 
\newcommand{\bbibitem}[1]{\bibitem{#1}\marginpar{#1}}
\newcommand{\llabel}[1]{\label{#1}\marginpar{#1}}

\def\Label#1{\label{#1}%
  \smash{\hbox to0pt{\raise1ex\hbox{\tiny[#1]}\hss}}}
\def\noLabels{\let\Label=\label}
\def\nobbibitem{\let\bbibitem=\bibitem}
\def\nolabell{\let\labell=\label}
\def\nollabel{\let\llabel=\label}

\title{{\bf
Typicality, Black Hole Microstates and Superconformal Field Theories
}}

\author{Vijay
Balasubramanian$^{1,2}$, Bart{\l}omiej Czech$^1$, Yang-Hui He$^{3}$,\\ Klaus Larjo$^1$ and Joan Sim\'{o}n$^4$ \footnote{vijay@physics.upenn.edu,
 czech@sas.upenn.edu,  hey@maths.ox.ac.uk,
 klarjo@physics.upenn.edu, \newline
J.Simon@ed.ac.uk}
\\[1mm]
\small \sl $^1$\; David Rittenhouse Laboratories, University of
Pennsylvania,
\\[-1.5mm]
\small \sl Philadelphia, PA 19104, USA \\
\small \sl $^2$ \;  School of Natural Sciences, Institute for Advanced Study, \\
[-1.5mm]
\small \sl  Princeton, NJ 08540, USA\\
\small \sl $^3$\; Merton College, OX1 4JD;
Mathematical Institute, OX1 3LB, \& \\
[-1.5mm]
\small \sl Rudolf Peierls Centre for Theoretical Physics, OX1 3NP, \\
[-1.5mm]
\small \sl University of Oxford, UK\\
\small \sl $^4$ \; School of Mathematics and Maxwell
Institute of Mathematical Sciences,
\\[-1.5mm]
\small \sl King's Buildings, Edinburgh, EH9 3JZ, UK  \\
}

\begin{document}

\noLabels
\nobbibitem
\nolabell
\nollabel
\setlength{\baselineskip}{16pt}
\begin{titlepage}
%
\maketitle

\begin{abstract} We analyze the structure of  heavy multitrace
BPS operators in $\mathcal{N} = 1$ superconformal  quiver gauge
theories that arise on the worldvolume of  D3-branes on an affine
toric cone.     We exhibit a geometric procedure for counting
heavy mesonic operators with given $U(1)$ charges.   We show that
for any fixed linear combination of the $U(1)$ charges, the
entropy is maximized when the charges are in certain ratios.  This
selects preferred directions in the charge space that can be
determined with the help of a piece of string.  We show that
almost all heavy mesonic operators of fixed $U(1)$ charges share a
universal structure.    This universality reflects the properties
of the dual extremal black holes whose microstates they create.   We also interpret
our results in terms of typical configurations of dual giant gravitons in AdS space.
\end{abstract}

\thispagestyle{empty} \setcounter{page}{0}
\end{titlepage}

\tableofcontents

\section{Introduction}

Superconformal field theories (SCFTs) living on the world-volume
of D3-branes transverse to the conical singularity of an affine
cone over a Sasaki-Einstein manifold $X$ are holographically dual
to gravity on ${\rm AdS}_5\times X$.   In this paper we explore the structure of heavy
BPS states in these theories,
describing their universal structural properties, and relating
them to dual extremal black holes in  ${\rm AdS}_5\times X$.

In \cite{plethystic}, generating functions for enumerating mesonic
BPS operators in these SCFTs were derived. This counting was
extended to include baryonic operators in \cite{giants,baryon},
but in the present paper we consider only zero baryon charge. The
authors of \cite{plethystic} derived the entropy associated to
mesonic BPS states carrying a fixed linear combination of the
three $U(1)$ charges. The most common and physically relevant
example is states of constant R-charge $R$.  In this article, we
focus on toric cones and introduce a geometric construction that
allows us to refine this counting and compute the entropy
associated to any triple of charges rather than a fixed linear
combination.  By extension we are able to single out the
particular triple that maximizes the entropy subject to any linear
constraint on the charges.   In particular this triple can be
computed by finding the center of gravity of the pyramid cut of
the dual toric cone by the constraint plane. Amusingly, this
amounts to suspending the pyramid from a piece of string,
providing a novel use for a different kind of string theory.

We will show that most heavy mesonic operators that are BPS have a
universal structure.  For a given set of $U(1)$ charges, we find
the mean distribution of trace factors in heavy multi-trace
operators.  We then quantify the scale at which this distribution
may be thought of as defining  a typical structure for heavy
operators, and spell out those features that are  shared by almost
all states. The heavy operators have gravity duals that can be
interpreted in terms of giant gravitons \cite{giants} or in terms
of dual giant gravitons \cite{sparks-Z}. This allows us to
reinterpret our results as defining  typical configurations of
giant gravitons and dual giants.   We also analyze the existence
of horizons in the dual supergravity solutions.

The organization of the paper is as follows.
In Section~\ref{sec:review} we briefly review the connection
between toric quiver theories and toric varieties. In Section
\ref{sec:entropy} we present our geometric construction and
compute the maximal entropy vector for any fixed linear
combination of the $U(1)$ charges. We also show that the entropies
found are not large enough to produce macroscopic horizons in the
dual gravity description.

In Section~\ref{sec:structure-dist}, we treat the system statistically, and find the mean distribution of traces for the
multitrace operators carrying any specified triplet of $U(1)$ charges.  We also give the interpretation in terms of dual giant
gravitons in $\ads{5}$.   Strictly speaking, since giant gravitons  and dual giant gravitons are compact objects,  their
wavefunctions can spread over their moduli spaces.  As we are analyzing these systems in terms of their typical classical
configuration, it is important to evaluate how well the wavefunction localizes at a given point in the configuration space.  For
the special case of $1/8$ BPS states in ${\cal N} = 4$ Yang-Mills theory, this analysis is carried out in
Appendix~\ref{typicalgiant}.    In Section~\ref{sec:gravity} we analyze the dual gravity description, and finally close in
Section \ref{discussion} with a discussion.  In Appendix~\ref{ap:amoeba} we point out some intriguing relations to topological
strings and amoebae.

\section{Review}\setall
\llabel{sec:review}

The AdS/CFT correspondence has been extended to a set of type IIB string backgrounds of the form $\textrm{AdS}_5\times X$, where
$X$ is a five-dimensional Sasaki--Einstein manifold \cite{sasakis}. The superconformal field theories dual to string theory on
these backgrounds arise from a stack of D3-branes located at the conical singularity associated to the metric cone of $X$, and
are 
quiver gauge theories. The material of the present paper applies to the most extensively studied class of the
correspondence - when the cone over $X$ is so-called {\bf toric}.

Recently, there has been a realization that toric geometry, quiver
theories, and dimer models give rise to an intimate web of
relations. In this section we briefly review these connections;
for a more comprehensive review the reader is referred to
\cite{krisreview}. With these prerequisites, we devise a geometric
construction in the dual toric cone to analyze BPS states carrying specified $U(1)$ charges.     A well-known example of a toric Calabi-Yau cone is the
complex cone over a complex surface known as the {\bf first del
Pezzo} surface \cite{toric}. Geometrically, this is the projective
plane $\mathbb{C}\mathbb{P}^2$ blown up at a point by a sphere. We
shall refer to this Calabi-Yau cone as $dP_1$ and shall use this
variety to illustrate this section and the ensuing discussions.

\subsection{Toric quiver theories and planar quivers}
\begin{figure}[t]
\begin{center}
\includegraphics[scale=0.4]{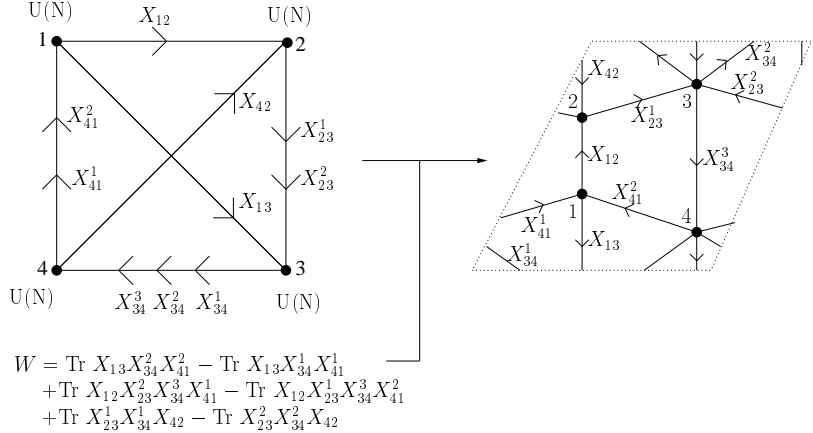}
\caption{Left: The quiver diagram and the superpotential
corresponding to $dP_1$. Right: The planar
quiver}\label{fig:DPjoint}
\end{center}
\end{figure}

\llabel{planar}
A toric quiver theory consists of a set of gauge
groups, $U(N_i)$ or $SU(N_i)$; for simplicity we take the groups
to have equal rank: $N_i \equiv N$ for all $i$. Further, the
theory contains a set of chiral superfields $X_j$, each transforming in
the bifundamental representation $(N,\bar{N})$ under a pair of
gauge groups and trivially under the others. The gauge groups and
the matter can be presented in a directed graph, with gauge groups
as nodes and matter multiplets as directed arrows between the two
nodes under which the multiplet transforms non-trivially. The
dynamics of the theory is encoded in a superpotential $W$. In
toric quiver theories the form of $W$ is highly restricted: it
consists of a sum of monomials, with each field appearing
linearly, and in exactly two terms with opposite signs. In Figure
\ref{fig:DPjoint} we portray the quiver diagram and the
superpotential of the quiver gauge theory corresponding to $dP_1$.

The toric quiver data and the superpotential can be incorporated
into a single graph called the planar quiver.  This is done by
embedding the quiver into a torus, and ``opening up'' the arrows
between the gauge groups in such a way that each arrow in the
planar quiver corresponds to exactly one field. This can always be
done so as to ensure that each term in the superpotential
corresponds to a face in the planar quiver, i.e., for any face the
arrows (fields) surrounding it make up a term in the
superpotential. Each face is circled by the fields either in a
clockwise or a counter-clockwise direction; the first corresponds
to a negative and the latter to a positive superpotential term.
See Figure~\ref{fig:DPjoint} for the planar quiver of $dP_1$.

\subsubsection{Mesonic operators in quiver theories}
We are interested in multi-trace mesonic gauge invariant
operators, which are of the form
\begin{equation}
O = \prod_{i=1}^k \textrm{Tr} \left( X_{b_{i1}}\,X_{b_{i2}}\,
\ldots\,X_{b_{ia_i}} \right). \Label{eq:operator}
\end{equation}
They form closed loops in the quiver and the planar quiver, with
definite winding numbers $(p,q)$ on the torus. The minimal mesonic
operators are loops that pass through any node of the quiver at
most once, i.e., operators that cannot be split into smaller
components by adding traces. These generate the chiral ring of
mesonic operators. Thus, any mesonic operator can be written as a
product of these minimal loops \emph{with suitably placed traces}.

The number of minimal operators can be computed from the quiver,
but not all of these operators are independent: we must impose the
vanishing of the F-terms coming from the superpotential $W$.
Therefore, the minimal operators are split into equivalence
classes under the F-flatness conditions $\partial W / \partial X_i
= 0$.   These equivalence classes are in
one-to-one correspondence with the winding numbers  on the torus, i.e., two
operators are equivalent if and only if they share the same
winding numbers $(p,q)$.

Applying this to our example, from the $dP_1$ quiver of Figure
\ref{fig:DPjoint} one can compute that there are 24 minimal loops,
and in Table~\ref{tab:DPloops} we split these loops into 9
equivalence classes. Verifying that these are the equivalence
classes under the F-term constraints is left as an exercise to the
reader.

\begin{table}[ht]
\begin{center}
\begin{tabular}{|c|c|c|}
\hline \# & Minimal loops & Winding $(p,q)$ \\
\hline \hline 1& $X_{12}X_{23}^2X_{34}^1X_{41}^2$ & $(-2,1)$ \\
\hline 3 & $X_{12}X_{23}^1X_{34}^1X_{41}^2, \quad X_{12}X_{23}^2X_{34}^1X_{41}^1, \quad X_{12}X_{23}^2X_{34}^2X_{41}^2$ & $(-1,1)$ \\
\hline 3 & $X_{12}X_{23}^2X_{34}^3X_{41}^2, \quad X_{23}^2X_{34}^1X_{42}, \quad X_{13}X_{34}^1X_{41}^2$ & $(-1,0)$ \\
\hline 3 & $X_{12}X_{23}^1X_{34}^1X_{41}^1, \quad X_{12}X_{23}^1X_{34}^2X_{41}^2, \quad X_{12}X_{23}^2X_{34}^2X_{41}^1$ & $(0,1)$ \\
\hline 6 & $X_{12}X_{23}^1X_{34}^3X_{41}^2, \quad X_{12}X_{23}^2X_{34}^3X_{41}^1, \quad X_{23}^2X_{34}^2X_{42}$ & $(0,0)$ \\ &
$X_{23}^1X_{34}^1X_{42}, \quad X_{13}X_{34}^1X_{41}^1, \quad X_{13}X_{34}^2X_{41}^2$ & \\
\hline 2 & $X_{13}X_{34}^3X_{41}^2, \quad X_{34}^3X_{42}X_{23}^2$ & $(0,-1)$ \\
\hline 1 & $X_{12}X_{23}^1X_{34}^2X_{41}^1$ & $(1,1)$ \\
\hline 3 & $X_{12}X_{23}^1X_{34}^3X_{41}^1, \quad X_{23}^1X_{34}^2X_{42}, \quad X_{13}X_{34}^2X_{41}^1$ & $(1,0)$ \\
\hline 2 & $X_{13}X_{34}^3X_{41}^1, \quad X_{23}^1X_{34}^3X_{42}$ & $(1,-1)$ \\
\hline
\end{tabular}
\end{center}
\caption{Equivalence classes of minimal loops and their windings
for the $dP_1$ quiver.} \label{tab:DPloops}
\end{table}

\subsection{Brane tilings and dimer models}
\Label{sec:dimers}

\begin{figure}[t]
\begin{center}
\includegraphics[scale=0.4]{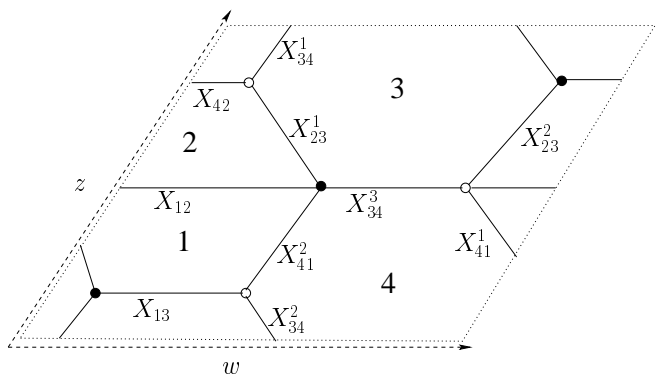}
\caption{The brane tiling (dimer model) for $dP_1$.}\label{fig:DPtiling}
\end{center}
\end{figure}

We can equivalently consider the graph dual to the planar quiver: this is obtained by replacing each vertex with a face, and each
arrow with a perpendicular line. This dual graph is called a {\bf brane tiling}. It has a natural bipartite structure, otherwise
known as a {\bf dimer model} \cite{krisreview}. If we color the nodes that correspond to negative superpotential terms in black,
and those corresponding to positive ones in white, we see that each vertex is only connected to vertices of opposite color. See
the brane tiling of $dP_1$ as an explicit example in Figure \ref{fig:DPtiling}; note that we have marked the two nontrivial
cycles on the torus and called them $z$ and $w$.

The advantage of working with the brane tiling is that it allows
us to find the toric Calabi--Yau manifold on the gravity side that
is dual to the quiver gauge theory. This is accomplished using the
Kasteleyn matrix $K$, which is formed from the brane tiling as
follows.

First, we mark some edges in the tiling with minus-signs. This is
done in such a way that for every face with (0 mod 4) edges
surrounding it, an odd number of those faces have minus-signs; and
for every face with (2 mod 4) surrounding edges, an even number
have minus signs. This can always be done for any brane tiling.
Then define the matrix elements $K_{ij}$, where the index $i$ runs
over the black nodes of the tiling, and $j$ runs over the white
nodes. $K_{ij}$ is a sum over all the edges connecting the two
nodes, with the following weights: -1, if the edge was marked with
a minus sign above; $z$, if the edge crosses the cycle $z$ in the
tiling in the positive direction; $z^{-1}$ if the edge crosses $z$
in the negative direction, and likewise for $w$ and $w^{-1}$. The
positive direction when crossing the cycles $z$ and $w$ can be
chosen arbitrarily, as long as it is consistently followed for all
elements.

This is best illustrated by writing down the Kasteleyn matrix
corresponding to the $dP_1$ brane tiling of Figure
\ref{fig:DPtiling}. In this case the minus-sign condition is
satisfied if we choose the edge corresponding to the field
$X_{12}$ to come with a $(-1)$.  Then, denoting the black vertices
1,2,3 from left to right and denoting the white vertices 1,2,3
from bottom to top, we can write down the Kasteleyn matrix:
\begin{equation}
K^{dP_1} (z,w) = \left( \begin{array}{ccc} 1 & z & w^{-1} \\ 1 & 1 - z & 1 \\ w & 1 & z^{-1}   \end{array} \right).
\end{equation}

\subsection{Characteristic polynomial and the toric cone}
\Label{sec:toricfan} Rather than $K_{ij}$ itself, the object we
really need is the determinant of $K$, which is the characteristic
polynomial $P(z,w)$. Since $P(z,w)$ is a polynomial in two
variables, each monomial $z^aw^b$ gives a point $(a,b)$ in a
two-dimensional lattice. These points make up a convex polygon
called the {\bf Newton polygon}, which in turn  determines a toric
variety. For details on the map between the toric variety and the
toric cone we refer the reader to \cite{toricreviews}. This
completes the circle of correspondences. The beautiful fact is
that this toric variety is precisely the singular Calabi-Yau cone
to whose tip the stack of D3-branes is transverse by construction.

\begin{figure}[t]
\begin{center}
\includegraphics[scale=0.4]{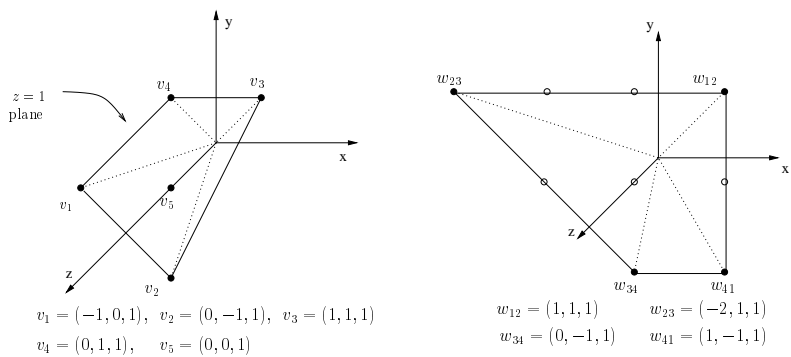}
\caption{Left: The toric cone of $dP_1$. Right: The dual cone. The
filled-in circles are the edge generators normal to the faces of
the toric cone, and the empty circles are the remaining
generators. Note that for this example the generators of the toric
cone lie on a plane; this does not hold for a general toric
variety, but only for Calabi-Yau ones.}\label{fig:toric}
\end{center}
\end{figure}

We embed the Newton polygon into the $z=1$ plane\footnote{~There
is nothing special about the $z=1$ plane: one can show that for a
toric Calabi--Yau manifold the endpoints of the generators of the
toric cone lie in a 2-plane, and without changing the variety one
can perform a $SL(3,\mathbb{Z})$ transformation to map this plane
to the $z=1$ plane. Also note our abuse of notation: we label the
coordinates on $\mathbb{Z}^3$ by $(x,y,z)$, but this $z$ is not
related to the dummy variable $z$ in the Kasteleyn matrix. As we
will not utilize the Kasteleyn matrix beyond this point, we hope
this does not cause any confusion.} in $\mathbb{Z}^3$, and denote
the boundary points of the $k$-gon by
$\vec{v}_0,\ldots,\vec{v}_{k-1}$, ordered anti-clockwise around
the polygon. Set $\vec{v}_k \equiv \vec{v}_0$. Linear combinations
of the vectors $\vec{v}_i$ form the toric cone $\mathcal{C}$ of
the variety; for details see again for instance
\cite{toricreviews}. For our long-running example, the del Pezzo
quiver, the characteristic polynomial is
\begin{equation}
P^{dP_1}(z,w) = \textrm{det } K_{ij}^{dP_1} = -4 + z^{-1} + w^{-1}+ z + zw,
\end{equation}
and the toric cone is shown in Figure~\ref{fig:toric}.

In addition to the toric cone $\mathcal{C}$, we also need to work with its
dual cone $\mathcal{C}^*$. The edges of the dual toric cone are
generated by the interior pointing normal vectors to the exterior
faces of the toric cone, normalized so that their magnitude is the
smallest that can be achieved with integer components. In other
words,
\begin{equation}
\vec{w}_{i,i+1} \propto \vec{v}_i \times \vec{v}_{i+1}.
\end{equation}
To find the other generators of the dual toric cone, one simply
needs to include the minimal set of interior points so that all
the integral points in the dual toric cone are spanned by the
generators. This is most easily illustrated by an example: in
Figure~\ref{fig:toric} we plot the generators of $\mathcal{C}^*$
for the del Pezzo quiver, with filled circles for the edge
generators, and empty circles for the interior points needed to
complete the set of generators of the dual cone.   The crucial
fact that enables the counting of operators from toric data is
that every integer point in the dual toric cone corresponds to a
unique single-trace BPS operator \cite{plethystic}.

\begin{figure}[t]
\begin{center}
\includegraphics[scale=0.4]{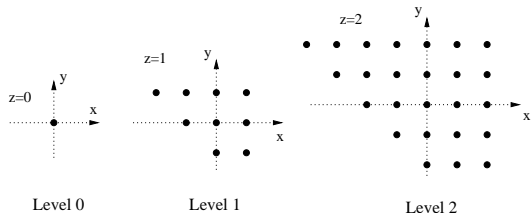}
\caption{The first three levels of the dual toric cone of $dP_1$, i.e., the slices $z=0,1,2$.
Here the level counts the number of minimal loops.
}\label{fig:DPlevels}
\end{center}
\end{figure}

Returning to our example, in Figure~\ref{fig:DPlevels} we show the
first three levels of points for the del Pezzo quiver. The single
point on level 0 is naturally the identity operator. On level 1
the reader should compare the $(x,y)$-coordinates of the points
with the winding numbers of operators in Table~\ref{tab:DPloops};
they match perfectly. Clearly the higher levels will match as
well, due to the conic structure of the construction.

Let us still flesh out this correspondence between points in the
cone and operators by writing down the operator corresponding to
point $(x,y,z)$.  This operator is given simply by
\begin{equation}
(x,y,z) \quad \leftrightarrow \quad A_{i_1} \ldots A_{i_k}, \quad \textrm{with } w_{i_1} + \ldots + w_{i_k} = (x,y,z),
\Label{correspondence}
\end{equation}
where the  $A$'s are the independent minimal loops defined in
Section~\ref{planar}, and $w_i$ is the generator of the dual toric
cone corresponding to the minimal loop $A_i$.  Thus the
correspondence above states that to construct the operator
corresponding to the point $(x,y,z)$, one has to choose a set of
generators (minimal loops) such that the sum of these generators
gives $(x,y,z)$.  There are generally many ways of choosing the
loops so that this condition is satisfied, but the beauty of the
construction is that all these operators will belong to the same
equivalence class under the F-term constraints, which makes it
possible to say that each point corresponds to a unique operator
in the quiver.

\subsubsection{$Z$-minimization and the Reeb vector}
\llabel{zminreeb} The final element we need is the Reeb vector
$\vec{b}$ of the toric variety. Its significance is that it
generates an isometry in the bulk geometry  corresponding to
the R-charge of the brane world-volume SCFT. Thus, in the dual
toric cone, the Reeb vector is normal to planes containing
operators of constant $R$. The directions generating an equal
R-charge plane are then the remaining flavor charges
\cite{krisreview}, defined up to a residual $SL(2,\mathbb{Z})$. As
shown in \cite{sparks-reeb}, the Reeb vector
$\vec{b}=(b_1,\,b_2,\,3)$ minimizes the volume cut from
$\mathcal{C}^*$ by a plane $2(\vec{b},\,\vec{n})=1$. This is
equivalent to minimizing the quantity $Z$:
\begin{equation}
Z(b_1,b_2) \equiv \sum_a
\frac{(\vec{v}_{a-1},\vec{v}_a,\vec{v}_{a+1})}
{(\vec{b},\vec{v}_{a-1},\vec{v}_a)(\vec{b},\vec{v}_a,\vec{v}_{a+1})}\,,
\Label{zminim}
\end{equation}
where $(\vec{v}_i,\vec{v}_j,\vec{v}_k)$ is the $3\times 3$
determinant constructed from the three column vectors. $Z$
essentially corresponds to the Einstein--Hilbert action for a
metric $h$ on $X$, and thus minimizing it, i.e., finding the
critical points, corresponds to finding the Sasaki--Einstein
metrics for this action. In the above we set $b_3=3$, which holds
in the $SL(3,\mathbb{Z})$ frame in which the Newton polygon is
embedded in the plane $z=1$. Thus finding the Reeb vector is
reduced to an algebraic problem of minimizing a function of two
variables. For $dP_1$, the minimization yields the Reeb vector
\begin{equation}
b^{dP_1} = (0,4-\sqrt{13},3). \Label{eq:DPReeb}
\end{equation}
If we assemble the three $U(1)$ charges of the BPS operator into a
vector $\vec{n} \in \mathcal{C}^*$, then the value of the R-charge
(and conformal dimension $\Delta$) is given by its scalar product
with the Reeb vector:
\begin{equation}
\Delta(\vec{n})=\frac{3}{2}R(\vec{n})=(\vec{b},\,\vec{n})\,.
\end{equation}
Exemplary values of the R-charge for integer points in the dual
toric cone of $dP_1$ are tabulated below. Notice that in general
the R-charge of an operator need not be integer-quantized or even
rational \cite{sparks-R}.

\begin{table}[t]
\begin{center}
\begin{tabular}{|c|c|c|}
\hline Level & Vector & Conformal Dimension \\ \hline
\hline 0 & (0,0,0) & 0 \\
\hline 1 & (a,1,1) & $7-\sqrt{13}$ \\
1 & (a,0,1) & 3 \\
1 & (a,-1,1) & $-1 + \sqrt{13}$ \\
\hline 2 & (a,2,2) & $2(7-\sqrt{13})$ \\
2 & (a,1,2) & $10-\sqrt{13}$ \\
2 & (a,0,2) & 6 \\
2 & (a,-1,2) & $2+\sqrt{13}$ \\
2 & (a,-2,2) & $2(-1+\sqrt{13})$ \\
\hline
\end{tabular}
\label{tab:Rs}
\end{center}
\caption{Conformal dimensions of charge vectors $\vec{n}$ in the dual toric cone of $dP_1$.
The level denotes the number of minimal loops. The R-charge is independent
of the first coordinate of $\vec{n}$ because $b_1=0$.}
\end{table}

\subsection{Toric cone as the phase space of dual giant gravitons}
\llabel{sec:dualphases}

So far we have concentrated on the description of the half-BPS
spectrum of mesonic scalar chiral primary operators
$\eH_{\text{mesonic}}$ in $\CN=1$ SCFT quiver gauge theories. Our
starting point was the mathematical description of toric cones
$\eM$ over Sasaki--Einstein manifolds $X$ probed by D3-branes.
However, the half-BPS sector of the gauge theory Hilbert space may
also be analyzed in terms of (dual) giant gravitons. In this
analysis, the machinery of toric geometry is as useful as it is on
the gauge theory side.

We recall that BPS mesonic sectors in SCFTs allow different
physical interpretations depending on the amount of R-charge
carried by the states. For states carrying order $N^0$ charge,
these are interpreted as pointlike rotating gravitons; when the
charge becomes of order $N$, they can also describe giant
gravitons or dual giants. Classically, the first set corresponds
to rotating D3-branes wrapping homologically trivial 3-cycles in
the cone $\eM$ over the base $X$; the second set describes
rotating D3-branes wrapping a 3-sphere in $\textrm{AdS}_5$. This
last approach was adopted in \cite{sparks-Z}, where the authors
showed that $\eM$ has a natural interpretation as the phase space
of the dual giant.

Using geometric quantization, the Hilbert space for the dual giant
$\mathcal{H}_{dg}$ was found to be given by the space of
holomorphic normalizable functions on $\eM$. Since $\eM$ is toric,
it was further shown that this space is spanned by the elements of
the dual toric cone $\mathcal{C}^*$. Thus, there exists an
isomorphism between $\mathcal{H}_{dg}$ and the dual toric cone
$\mathcal{C}^*$.

Since dual giants are mutually BPS, the $N$-dual giant states can
be described by $N$ indistinguishable quantum particle states.
Hence the $N$-dual giant Hilbert space is simply the $N$-th
symmetric tensor product
\begin{equation}
\mathcal{H} = {\rm Sym}^N \mathcal{H}_{dg}\,.
\end{equation}
This fact allows to compute the partition functions of these systems.
We will be interested in the grand canonical partition function for multi-dual
states, which was found in \cite{sparks-Z} and agrees with the
grand canonical partition function of the CFT presented in
\cite{plethystic}. We will be analyzing these partition functions
in Section~\ref{sec:structure-dist}, and interpreting the results
both in terms of typical mesonic CFT operators, and in terms of
typical configurations of dual giant gravitons.

\paragraph{Isomorphism of Hilbert spaces:} Besides the isomorphism of Hilbert spaces described above, there exists a further one between the Hilbert space of giant gravitons and $\mathcal{H}$  \cite{giants}. This can be seen as follows.

The classical moduli space of giant gravitons is given by the set of polynomials of degree $N$ in
$\eM$ subject to a set of constraints \cite{giants}
\begin{equation}
  P(z_1,\,z_2,\,z_3) = c+c_iz_i+c_{ij}z_iz_j+\dots = \sum_I c_Iz_I\,.
\end{equation}
Following \cite{beasley}, the quantum Hilbert space for
such giants ${\cal H}_g$ is spanned by the states
\begin{equation}
  |c_{I_1},\,c_{I_2},\,\dots ,c_{I_N}\rangle \,.
\end{equation}
These states are holomorphic polynomials of degree $N$ over the
classical moduli space and are symmetric in the $\{c_{I_i}\}$.
Thus, ${\cal H}_g$ is the symmetric product
\begin{equation}
  {\cal H}_{\text{g}} = \text{Sym}\left(|c_{I_1}\rangle \otimes |c_{I_2}\rangle \otimes \dots
  \otimes |c_{I_N}\rangle
 \right)\,,
\end{equation}
where each $|c_{I_i}\rangle$ represents an holomorphic function over the cone $\eM$ with base $X$.
This establishes the isomorphism between the Hilbert spaces of giant gravitons and dual giants.

We want to emphasize that this last isomorphism is captured by the
combinatorial identity, a cornerstone of the plethystic program:
\begin{equation}
  g_N(t,\,\eM) = g_1(t,\,\eM^N/S_N)\,,
 \label{fund-pleth}
\end{equation}
where $g_N(t,\,\eM)$ stands for the partition function of the
gauge theory of $N$ D3-branes probing $\eM$, whereas
$g_1(t,\,\eM^N/S_N)$ stands for the partition function of a single
D3-brane in the symmetric product  $\eM^N/S_N$.

The full discussion is summarized by the equality :
\begin{equation}
  \eH_{\text{g}} = {\rm Sym}^N \eH_{dg} = \eH_{\text{mesonic}}\,.
\end{equation}
We conclude that the partition function $g_N(t,\,\eM)$ counts
half-BPS mesonic scalar chiral primary operators, \emph{and} giant
graviton states carrying a certain R-charge, \emph{and} dual giant
graviton states carrying the same R-charge.

\section{Counting multi-trace operators}\setall
\llabel{sec:entropy}

In the previous section we learned how each point $\vec{n}$ in the
dual toric cone corresponds to an operator in the quiver theory
via (\ref{correspondence}). These operators are not gauge
invariant as each field transforms in a non-trivial representation
under two of the gauge groups. In order to create gauge invariant
operators we need to contract appropriately the indices on the
fields $X_i$. This can be done by inserting traces at appropriate
places in the operator.\footnote{~By constructing gauge invariant
operators using traces we are picking a preferred basis that is
convenient for us and which is related to multi-particle states in
AdS space. Another possible choice for constructing such operators
would be to use determinants and sub-determinants which are
related to D-branes in AdS space.}

The simplest way of making the operator in (\ref{correspondence})
gauge invariant is to take the trace of the entire quantity. This
leads to a single-trace operator, which corresponds canonically to
a point in the dual toric cone. As mentioned below
eq.~(\ref{correspondence}), this correspondence is bijective and
extends to all the integer points in $\mathcal{C}^*$. Therefore
every integer-coordinate point in the dual cone is uniquely
associated with a single-trace gauge invariant operator. However,
we can also create a gauge invariant operator by distributing
several traces in the product (\ref{correspondence}) to produce
multi-trace operators of the form (\ref{eq:operator}). This may
clearly be done in a multitude of ways. In this section we will
compute the number of such inequivalent trace structures. This is
important because for large charges the corresponding field theory
states are expected to be dual to black hole microstates, and our
computation of the degeneracy of operators for any triplet of
charges provides a prediction for the entropy of these black
holes. We return to this in more detail in Section
\ref{sec:gravity}.

Before proceeding, we should explain how our counting differs from
the one done in the plethystic program \cite{plethystic,baryon}.
There, the authors computed the entropy of mesonic operators of a
given R-charge and provided an implicit expression--the refined
plethystic exponential--counting operators at any given
\emph{point} in the dual cone (meaning operators of a given
R-charge and flavor charges $n_1$ and $n_2$). In this paper, we
pursue a different strategy, which involves introducing a new
slicing of the dual toric cone. This method affords an elegant
geometric interpretation, and, crucially, allows us to find the
\emph{asymptotics} of the {\it refined} counting, which was not
done in the plethystic program. As one expects the flavor charges
of black holes to be measurable to classical observers, it is the
refined counting carried out in this paper that is expected to
match the entropies of the dual black holes.

There is a simple way of stating our counting problem: since each
trace component in the expression (\ref{eq:operator}) is a
single-trace operator, it corresponds to a point in the dual toric
cone. Thus, computing the number of ways of distributing traces is
equivalent to finding the number of partitions of the vector
$\vec{n}$ into components: $\vec{n} = \sum_j \vec{m}_j$. The
asymptotics of this type of counting problems can be tracked using
a theorem due to Meinardus \cite{meinardus}; our application of
Meinardus' theorem will follow that of \cite{plethystic}. We shall
proceed in the following steps: First, we count the number of
partitions for all vectors on a fixed plane; this is equivalent to
counting multi-trace operators with a single fixed charge, say
R-charge. We then go on to count the vectors on this plane
individually; this is the refinement, wherein a particular triplet
of $U(1)$-charges is fixed. We carry out the computation in this
way because, as we shall see in the ensuing arguments, this method
leads to a handy geometric construction involving centers of
gravity of certain pyramids.

\subsection{Step 1: Partitioning co-planar vectors}
\llabel{partplanar}

As the first step we shall reproduce the non-refined plethystic counting,
and use the Meinardus theorem \cite{meinardus} to find the asymptotic
number of multi-trace operators on any plane of rational slope containing the
point $\vec{n}$ in the dual toric cone. Let us pick such a plane
and denote it by $p$. Then consider an auxiliary family $\{p_m\}$ of planes
parallel to $p$, such that (1) each point in the dual cone is
contained in one of the planes, and (2) the planes are evenly
placed, i.e., the distance between any two neighboring planes is
fixed.    In this auxiliary family we index the planes from the tip of the cone, and take $p_n \equiv p$ to be the plane that contains the point $\vec{n}$.

\begin{figure}[ht]
\begin{center}
\includegraphics[scale=0.60]{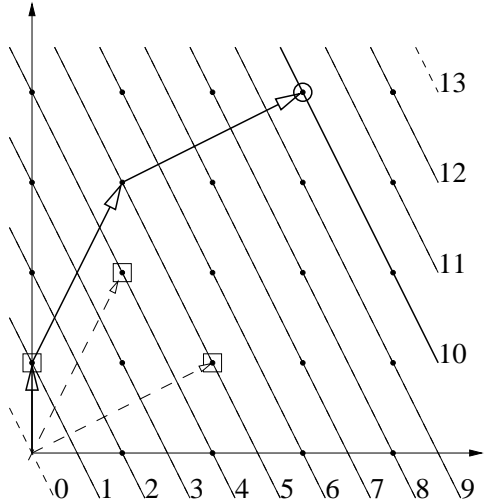}
\end{center}
\caption{A two-dimensional partition example with one possible family
of parallel lines with slope $-2$.  The dark lattice points
are the coordinate points and the numbers label the lines: $p_n$ means the
$n$-th line with slope $-2$ away from the origin.
We have specifically chosen to draw the line $p_{10}$ as a solid
line while other lines $p_m$ are dotted and for $p_{10}$,
we have portrayed one possible partition of vectors summing up to the
lattice points on it.}
\label{rain}
\end{figure}

To illustrate this, we display a two-dimensional analogue  in
Figure~\ref{rain}; here the cone is chosen to be the positive
quadrant of $\mathbb{Z}^2$, and parallel planes are drawn as
parallel lines. Next we wish to find the sum of the
 number of different vector partitions (which we will call the {\it multiplicity}) of  all the points on the plane $p_n$.  This is the
first step on the way to finding the multiplicity of the point
$\vec{n}$.  As is clear from the above setup and Figure
\ref{rain}, this problem is equivalent to enumerating the sets of
vectors $\{\vec{m}_j\}$ such that $\sum_j \vec{m}_j$ is on the
plane $p_n$.    Let $m_j$ index the plane that contains $\vec{m}_j$, counting from the tip of the cone.
Then this condition is equivalent to
\begin{equation}
\sum_j m_j = n.
\end{equation}
Let us denote the number of  integer coordinate points on the plane $p_m$ by $a_m$. Then the problem is equivalent to the problem of partitioning an
integer $n$ in integer components $m_j$, with the added complication that at level $m$ there are $a_m$ inequivalent summands.
This is a standard partition problem and the asymptotics is
studied by Meinardus \cite{meinardus}; his theorem states that
asymptotically the number of such partitions is given by
\begin{equation}
\label{meinardus1} S(n) \equiv \log \{\#\, {\rm partitions}\, {\rm of}\, n\} = \frac{\alpha + 1}{\alpha} \Big(A n^\alpha\,
\Gamma(\alpha + 1)\, \zeta(\alpha+1)\Big)^{\frac{1}{\alpha+1}} + \mathcal{O}(\log{n}) \, .
\end{equation}
Here $A$ is the residue at the rightmost pole, located at
$s=\alpha$, of the Dirichlet series $D(s)$ given by
\begin{equation}
D(s) \equiv \sum_{m=1}^\infty \frac{a_m}{m^s}\, . \Label{dirichlet}
\end{equation}
Therefore, to find the multiplicities we need to know the
coefficients $a_m$, i.e., the number of (integer coordinate)
points on plane $p_m$. In the plethystic program, the integer
$a_m$ corresponds to the number of single-trace operators at level
(distance) $m$, and the multiplicities we are after are integers
$d_j$ such that
\begin{equation}
\prod\limits_{m=1}^\infty (1-t^m)^{-a_m} =
\sum\limits_{j=0}^\infty d_j \,t^j\,.
\end{equation}
Here the exponents of $t$ parameterize the distance of a plane
parallel to $p$ from the tip of the dual cone. The quantities
$d_j$ are the numbers of multi-trace operators composed of the
single trace ones. We will revisit this in Section
\ref{sec:structure-dist}.

\paragraph{Example:}
Let us see how this works for the two dimensional case portrayed in Figure~\ref{rain}, where we have chosen the lines (planes) to
have slope $-2$ and the numbers $a_m$ are given by $a_m = 1,1,2,2,3,\ldots \to 1+[m/2]$. We see that indeed, $a_0, a_1, a_2,
\ldots$ correspond to the number of points on the lines $p_0, p_1, p_2, \ldots$ Clearly, for a general slope\footnote{~We assume
$a>1$; the opposite case clearly works similarly.} ($-a$) the coefficients are given by
\begin{equation}
a_m = 1 + \left[ \frac{m}{a} \right]\,\, \stackrel{m \gg
1}{\approx}\,\, \frac{m}{a} + \mathcal{O}(1),
\end{equation}
where $[x]$ denotes the integer part of $x$.

Plugging $a_m$ into (\ref{dirichlet}), we can evaluate the Dirichlet series to give
\begin{equation}
D(s) = \frac{1}{a} \zeta(s-1) + \cdots \, ,
\end{equation}
from which we find the location of the rightmost pole at $s=2$,
with residue $1/a$, i.e., $\alpha=2$ and $A=1/a$.   The omitted
terms give additional poles which are further to the left. Thus
the Meinardus theorem yields
\begin{equation}
\label{meinardusplane} S(p_n) = \frac{3}{2} \Big(\frac{n^2}{a}\, \Gamma(3)\, \zeta(3)\Big)^{\frac{1}{3}} + \mathcal{O}(\log{n})
\, .
\end{equation}
Note that the quantity $An^{\alpha} = n^2/a$ in the Meinardus
theorem is equal to twice the area of the triangle cut off from
the cone by the $p_n$ line. This will be true also in the three
dimensional case: the coefficient $An^{\alpha}$ will be
proportional to the volume of the pyramid cut off from the cone by
the plane $p_n$. This will be of critical importance later; it
allows us to sidestep intractable analytic computations via an
elegant geometric construction.

\subsubsection{Explicit enumeration of the planar partitions}
Let us now generalize from the two-dimensional example to a
$d$-dimensional toric variety. For us the interesting dimension is
$d=3$, but working in a general dimension comes at no additional
cost.

We start with the coefficients $a_m$: the number of integral
points of the dual toric cone contained in the plane $p_m$. For
large $m$, the number of integral points can be approximated by
the ($d-1$)-dimensional area of the plane $p_m$. By dimensional
arguments, this area scales as $a_m = \mathcal{A} m^{d-1}$, where
$\mathcal{A}$ is a constant specified by the orientation of the
plane in the cone. This allows us to evaluate the Dirichlet series
(\ref{dirichlet}) as
\begin{equation}
D(s) = \sum_{m=1}^\infty \frac{\mathcal{A}\, m^{d-1}}{m^s} +
\textrm{corrections} =\mathcal{A}\, \zeta(s+1-d) +
\textrm{corrections}\, .
\end{equation}
This has the rightmost pole at $s=d$, with residue $\mathcal{A}$,
yielding the parameters $\alpha=d$ and $A=\mathcal{A}$.

To evaluate $\mathcal{A}$, consider the number of integral points
inside the pyramid cut off from the dual toric cone by the plane
$p_m$. This can be approximated by the volume $V_m$ of the
pyramid. Again by dimensional arguments, for a $d$-dimensional
cone this volume scales as $V_m = \lambda m^d$ for some $\lambda$.
Clearly the number of points in the plane $p_m$ can then be
written as
\begin{equation}
a_m = V_m - V_{m-1} = \lambda (m^d - (m-1)^d) \approx \lambda d m^{d-1} + \mathcal{O}(m^{d-2}) \approx \frac{dV_m}{m},
\Label{eq:am}
\end{equation}
from which we find the residue
\begin{equation}
\mathcal{A} \equiv \frac{a_m}{m^{d-1}} \approx \frac{dV_m}{m^d}.
\end{equation}
Using these results we can plug into (\ref{meinardus1}) and write the number of partitions of vectors on plane $p$ as
\begin{equation}
S(p) \equiv \ln (\textrm{\# partitions}) = \frac{d + 1}{d}
\Big(Vd\, \Gamma(d + 1)\, \zeta(d+1)\Big)^{\frac{1}{d+1}} +
\mathcal{O}(\log{V}) \, , \Label{meinardus2}
\end{equation}
where $V$ is the volume of the pyramid cut off from the cone by
the plane $p$.   We have dropped the index $n$ in $p_n$ because it
is a function only of the particular plane in question, and not
the additional family of auxiliary planes used in the
construction.

We will need an explicit equation for the volume $V$ later in
Section~\ref{sec:structure-dist} and will give the somewhat
complicated expression there. For now we do not need such an
explicit formula; it would even be a hindrance, as the analytic
computations would quickly become hopelessly complicated, while
simple geometric arguments can take us remarkably far.

\subsection{Step 2: Partitions of individual vectors}
\llabel{partind}

Having found the partitions of all vectors lying on a given plane
$p$, we now wish to use the results above to find the possible
partitions of an individual vector $\vec{n}$. We shall accomplish
this by considering an upper bound on the number of such
partitions, and then showing that asymptotically this bound is
saturated.

Consider $\{p_{\vec{n}, \vec{s}}\}$, the set of all planes
containing the point $\vec{n}$. Here the subscript $\vec{s}$
indicates that the plane $p_{\vec{n}, \vec{s}}$ is defined as
being normal to $\vec{s}$, as well as containing $\vec{n}$. Note
that for $p_{\vec{n}, \vec{s}}$ to cut off a finite pyramid from
the cone, we must have $(\vec{s},\vec{w}_a) > 0$ for all the edge
generators $\vec{w}_a$ of the dual cone.\footnote{~Since $\vec{s}$
and $-\vec{s}$ define the same plane, $(\vec{s},\vec{w}_a) < 0$
would work just as well. We choose $>0$ without loss of
generality.} One can see that this condition is equivalent to
$\vec{s}$ living inside the toric cone.\footnote{~Note: not the
dual cone $\mathcal{C}^*$, but the toric cone $\mathcal{C}$.}
Continuing the above notation, we denote the volume of the pyramid
cut out by the plane $p_{\vec{n}, \vec{s}}$ as
$V_{\vec{n},\vec{s}}$.

Since the point $\vec{n}$ is merely a point in the plane
$p_{\vec{n},\vec{s}}$, the multiplicity of $\vec{n}$ is
bounded by the multiplicity of $p_{\vec{n},\vec{s}}$.
This gives:
\begin{equation}
e^{S(\vec{n})} \le e^{S(p_{\vec{n},\vec{s}})} = \exp \left(
\frac{d+1}{d} \left( d\, V_{\vec{n},\vec{s}}\, \Gamma(d+1)
\zeta(d+1) \right)^{\frac{1}{d+1}} \right) \, . \Label{estimate}
\end{equation}
Since this is true for any plane $p_{\vec{n},\vec{s}}$, we can
take a minimum over all the possible $\vec{s}$ on the right hand
side of (\ref{estimate}). This minimum is well defined because
$\vec{s}$ lives inside the toric cone, and on the faces of the
cone $V_{\vec{n},\vec{s}}$ diverges.\footnote{~This is because for
$\vec{s} \in \partial \mathcal{C}$, the constraining planes become
parallel to faces of $\mathcal{C}^*$. Thus, they fail to contain
finite volumes within $\mathcal{C}^*$ so $V_{\vec{n},\vec{s}}$
diverges.} Thus a minimum will exist inside the cone. Defining the
function $f(\vec{n})$ as
\begin{equation}
f(\vec{n}) = d \min_{\vec{s}} V_{\vec{n}, \vec{s}}, \Label{define-f}
\end{equation}
we can write (\ref{estimate}) as
\begin{equation}
e^{S(\vec{n})} \le \exp \left( \frac{d+1}{d} \left( f(\vec{n}) \Gamma(d+1) \zeta(d+1) \right)^{\frac{1}{d+1}} \right).
\Label{estimate2}
\end{equation}
Now we wish to argue that in an asymptotic limit this bound
becomes saturated, and therefore (\ref{estimate2}) gives us the
multiplicity we are after.

It is straightforward to see which $\vec{s}$ minimizes the
expression (\ref{define-f}): by construction the volume minimizing
plane,  $p_{\vec{n},\vec{s}_{min}}$, has to be tangent to surfaces
of constant $f$, and therefore\footnote{~To be precise, this
argument only implies that $\vec{s}_{min}$ is proportional to the
gradient $\vec{\nabla} f$, but since the normalization of
$\vec{s}$ is immaterial, we can use this equation to fix the
normalization.}
\begin{equation}
\vec{s}_{min} = \vec{\nabla} f(\vec{n}), \quad \textrm{so that }  f(\vec{n}) = d V_{\vec{n},\vec{\nabla}f(\vec{n})}.
\Label{fvol}
\end{equation}
Thus, to any point $\vec{n}$ we can associate the volume
minimizing plane
\be
p(\vec{n}) \equiv p_{\vec{n},\vec{s}_{min}} \, .
\ee
In Sec.~\ref{centerofgravity} below we show that $\vec{n}$ is the
center of gravity of the polygon $p(\vec{n}) \cap \mathcal{C}^*$.
The uniqueness of the center of gravity then establishes that
$p(\vec{n})$ is injective. But this implies that the point
$\vec{n}$ maximizes the multiplicities on the plane $p(\vec{n})$.
This is because, by the injectivity of the map $p$, all other points
$\vec{n}'$ on the plane $p(\vec{n})$ must have different
minimizing planes $p(\vec{n}')$ for which $\vec{n}'$ will be the
center of gravity.
Then, necessarily, $V_{p(\vec{n}')}
< V_{p(\vec{n})}$, since $p(\vec{n}')$ is volume-minimizing plane for
$\vec{n}'$. Thus,
\begin{equation}
e^{S(\vec{n}')} \le e^{S(p(\vec{n}'))} < e^{S(p(\vec{n}))} = \exp \left( \frac{d+1}{d} \left( f(\vec{n}) \Gamma(d+1) \zeta(d+1)
\right)^{\frac{1}{d+1}} \right). \Label{saddle}
\end{equation}
This then implies that
\begin{equation}
{e^{S(p(\vec{n}))} \over  A(p(\vec{n})) } \leq e^{S(\vec{n})} \leq e^{S(p(\vec{n}))}
\Label{multineq}
\end{equation}
where $A(p(\vec{n}))$ is the area of the volume minimizing plane
passing through $\vec{n}$. The left hand inequality simply uses
the fact that $S(\vec{n}) > S(\vec{n}^\prime)$ on the plane
$p(\vec{n})$  while $e^{S(p(\vec{n}))} = \sum_{\vec{n}^\prime}
e^{S(\vec{n}^\prime)}$.   The right hand inequality is
(\ref{estimate2}).  Now observe that the $A(p(\vec{n}))$ scales as
$|\vec{n}|^{d-1}$ where we are taking the Euclidean norm.  Thus,
taking logarithms of both sides of (\ref{multineq}) we get
\begin{equation}
S(p(\vec{n})) - \mathcal{O}(\log(|\vec{n}|)) \leq S(\vec{n}) \leq
S({p(\vec{n}))}
\end{equation}
Now from (\ref{meinardus2}) $S(p(\vec{n}))$ scales as $V^{{1 \over
d+1}} \sim |\vec{n}|^{{d \over d+1}}$. Thus we see that when
$|\vec{n}|$ is large,
\begin{eqnarray}
e^{S(\vec{n})} & \approx & e^{S(p(\vec{n}))} = \exp \left( \frac{d+1}{d} \left( f(\vec{n}) \Gamma(d+1) \zeta(d+1)
\right)^{\frac{1}{d+1}} \right), \quad {\rm or} \nonumber \\
S(\vec{n}) & \approx &  \frac{d+1}{d}  \left( f(\vec{n}) \Gamma(d+1) \zeta(d+1) \right)^{\frac{1}{d+1}} \,.
\Label{meinardus3}
\end{eqnarray}
This is the main result of this section.
We have only kept
leading terms in all expressions; to accurately evaluate
multiplicities near the boundaries of $\mathcal{C}^*$ subleading
terms will have to be kept too.   Eq.~(\ref{meinardus3}) can be
used to compute the number of multi-trace operators corresponding
to any given triplet of charges, and thus gives a prediction for
the entropy of the dual black holes on the gravity side.

\paragraph{$Z$-minimization revisited:}
Recall from Sec.~\ref{zminreeb} that the Reeb vector $\vec{b} =
(b_1, b_2, 3)$ is computed by minimizing $Z$, which is
proportional to the volume cut from $\mathcal{C}^*$ with a plane
of the form $2(\vec{b},\,\vec{n})=1$ \cite{sparks-Z}. Notice that all such planes
pass through the point $\vec{n} = (0,0,\frac{1}{6})$. The Reeb
vector is therefore precisely the normal to the plane
$p((0,0,\frac{1}{6}))$ and indeed to all planes $p(0,0,z)$:
\begin{equation}
\vec{b} \propto \vec{\nabla} f((0,0,z))\,, \Label{reebexplicit}
\end{equation}
where the proportionality is set by requiring that $b_3 = 3$. The
argument above eq.~(\ref{saddle}) then establishes that every
point on the $z$-axis is the maximal entropy point on its own
equal R-charge surface. Because these surfaces take the form
\begin{equation}
(\vec{b},\,\vec{n}) = b_1 x + b_2 y +3 z = \frac{3R}{2}\, ,
\end{equation}
the point of maximal entropy of an equal R-charge surface is
always given by
\begin{equation}
(0,0,\frac{R}{2})\, . \Label{maxR}
\end{equation}
In the next subsection we generalize this to arbitrary constraint
surfaces. The result (\ref{maxR}) holds in the standard
presentation of the dual toric cone, that is when the Newton
polygon lives in the plane $z=1$.

\begin{figure}[t]
\begin{center}
\includegraphics[scale=0.8]{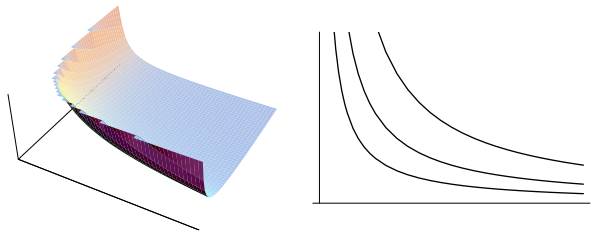}
\caption{Equal entropy surfaces in the dual toric cone of
$\mathbb{C}^3$. They take the form of generalized hyperboloids (in
the interior of the cone; see left) and hyperbolae (on the faces
of the cone; see right.) The interior surfaces represent
parametrically higher entropies. }\label{equalentropy}
\end{center}
\end{figure}

\paragraph{Surfaces of constant entropy:}
From (\ref{meinardus3}) we see that the surfaces of constant
entropy in the dual toric cone are given by level sets of the
function $f(\vec{n})$. From the definition of $f(\vec{n})$, these
surfaces are characterized by the property that at each point
$\vec{n}$ on the surface, the plane tangent to the surface at
$\vec{n}$ cuts out an equal volume from the cone. Surfaces with
this property are generalized hyperboloids in the cone. In
Figure~\ref{equalentropy} we plot these surfaces of constant
entropy.

It is also interesting to consider entropy on the faces of the
cone. Although the preceding construction appears to assign zero
entropy to these faces, this should taken as a statement that the
entropy on the faces is parametrically smaller than inside the
cone. A good example here is $\mathbb{C}^3$, where the interior of
the cone corresponds to the $\frac{1}{8}$-BPS sector, i.e.,
operators composed of three fields \{X,Y,Z\}, while on the faces
only two fields (say \{X,Y\}) are available. Evidently the entropy
on the faces is parametrically lower but non-zero. The equal
entropy curves on the faces take the form of hyperbolae. (See
Figure~\ref{equalentropy}.)

\subsection{Maximizing entropy}

In Section~\ref{partind} we saw that the degeneracy of multi-trace
operators with a specified $d$-tuple of charges $\vec{n}$ is
entirely determined by the behavior of one function, $f(\vec{n})$,
which was essentially given by the minimal volume that can be cut
out from the dual toric cone with a $(d-1)$-plane containing
$\vec{n}$. The normal to this minimizing plane is then given by
the gradient $\vec{\nabla} f(\vec{n})$, and it is very useful to
think of $\vec{\nabla} f(\vec{n})$ as the direction of the most
rapid growth of entropy, starting from the point $\vec{n}$.

A problem of interest is to maximize entropy subject to a
specified constraint, such as a fixed R-charge $R$. In the
following we consider constraints which are linear in the charge
vector $\vec{n}$. Such constraints trace ($d-1$)-hyperplanes in
$d$-dimensions and take the form
\begin{equation}
c_{\vec{s}}(\vec{n}) = (\vec{s}, \vec{n}) - n_s = 0,
\end{equation}
where $\vec{s}$ is the normal of the constraint plane, and $n_s$
is proportional to the corresponding charge, with the
proportionality constraint set by the normalization of $\vec{s}$.
For the special case of fixed R charge, the normal to the
constraint planes is given by the Reeb vector $\vec{b}$ of the
toric variety, and the constraint equation reads
$(\vec{b},\vec{n}) - \frac{3}{2} R = 0$.

Maximizing entropy given a constraint $c_{\vec{s}}(\vec{n})=0$ is then accomplished using the standard Lagrange multiplier
technique: the maximum is given by the point where the constraint plane is tangent to a surface of constant $f(\vec{n})$ which is
equivalent to
\begin{equation}
\vec{\nabla}f(\vec{n}) \propto \vec{\nabla} c_{\vec{s}}(\vec{n}) =
\vec{s}. \Label{lagrange}
\end{equation}
Unfortunately, due to the complicated analytic expression for the volume of a pyramid cut from the cone, equation
(\ref{lagrange}) is  difficult to work with analytically. Luckily, a simple geometric observation reduces this
problem to a very tractable one, familiar from freshman mechanics.

\subsubsection{Maximal entropy points as centers of gravity}
\llabel{centerofgravity}

We shall now show that the point of maximal multiplicity on a given plane is given by the center of gravity of that plane. This provides a simple prescription for finding the maximal entropy given an arbitrary
constraint.   Using the AdS/CFT correspondence, this will give the maximal entropy of a black hole in AdS space with charges satisfying the constraint.   In particular, this will allow us to find  the entropy and flavor charges of the most entropic black hole with R charge $R$.

\begin{figure}[t]
\begin{center}
\includegraphics[scale=0.4]{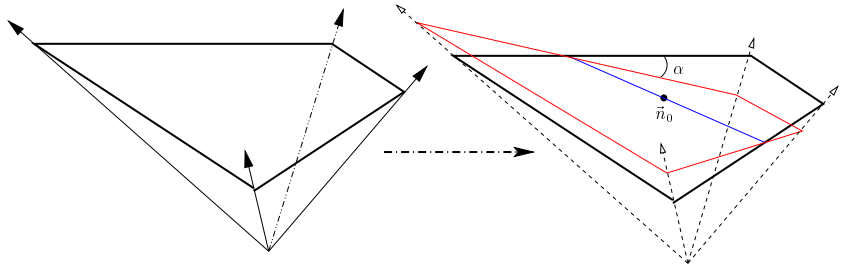}
\caption{Left: A constraint plane slicing the dual toric cone. Right: A line $l$ (in blue) through the center of gravity of the
slice. The contour in red illustrates the change in the volume cut by the plane after it is tilted about $l$ by the angle
$\alpha$. The toric variety used in the present example is the cone over $dP_1$.}\label{torquefig}
\end{center}
\end{figure}

Consider a linear constraint $c_{\vec{s}}(\vec{n}) = 0$, and
denote the constraint plane by $p_{\vec{s}}$. Let  the center of
gravity of $\,p_{\vec{s}} \cap \mathcal{C}^*$ be $\vec{n}_0$. Now,
pick an arbitrary line\footnote{~For clarity, we present this
argument for the three-dimensional cone. However, the same logic
will go go through in an arbitrary dimension $d$.} $l$ contained
in the plane $p_{\vec{s}}$ that passes through the center of
gravity $\vec{n}_0$. This is portrayed in Figure~\ref{torquefig}.
To find the center of gravity, we define coordinates on the plane
$p_{\vec{s}}$. We do so by choosing one of the coordinates, $x$,
to be along $l$, and the other, $y$, to be perpendicular to $l$.
Then the center of gravity is defined by the condition
\begin{equation}
\int_{A} dx \, dy\, y = 0\, , \Label{torque}
\end{equation}
where the integration is over the part of the plane $p_{\vec{s}}$
contained in the dual toric cone, i.e., $A = p_{\vec{s}} \cap
\mathcal{C}^*$.

Now, consider tilting the plane $p_{\vec{s}}$ around the axis $l$ by a small angle $\alpha$, as illustrated in Figure
(\ref{torquefig}). Denoting this tilted plane by $p_{\vec{s}}'$, the change in volume cut from the cone can be written as
\begin{equation}
\Delta V = \int_{A'} dx \, dy\, y \sin \alpha \approx \alpha\!
\int_A dx \, dy \, y + \mathcal{O}(\alpha^2), \Label{deltaV}
\end{equation}
where $A'$ is the intersection of the tilted plane $p_{\vec{s}}'$
and the dual cone $\mathcal{C}^*$. The approximation in the
equation above requires a little bit of explanation, as one needs
to take into account the change of region of integration from $A'$
to $A$. However, it is easy to see that this change will
contribute terms that are proportional to $\alpha$, and combined
with the $\sin \alpha$ factor these corrections only show up at
order $\mathcal{O}(\alpha^2)$. The condition that $p_{\vec{s}}$ is
the volume minimizing plane, $\Delta V = 0$, is therefore
equivalent to setting the torque about the axis $l$ to vanish as
in (\ref{torque}).

However, since the line $l$ is an arbitrary line passing through
the center of gravity $\vec{n}_0$, the above argument is valid for
any such line, and therefore for any tilt about $\vec{n}_0$. It
follows that the constraint plane $p_{\vec{s}}$ is the volume
minimizing plane for the center of gravity $\vec{n}_0$, and
therefore, the arguments from Section~\ref{partind} tell us that
$\vec{n}_0$ must be the most entropic point on the plane
$p_{\vec{s}}$.

As a side note, the preceding argument affords a new
characterization of the Reeb vector. In particular, $\vec{b}$ is
the normal to the unique family of parallel planes $p_n$ such that
the center of gravity of $p_n \cap \mathcal{C}^*$ is contained in
the $z$-axis ($x_d$-axis in $d$ dimensions).\footnote{~With a
slight abuse of notation, we use $z$ to denote a coordinate both
on the toric cone and the dual toric cone.} This observation is
applicable in the standard presentation of the toric fan of
$\mathcal{M}$, where the Newton polygon is contained in the plane
$z=1$.

\begin{figure}[t]
\begin{center}
\includegraphics[scale=0.4]{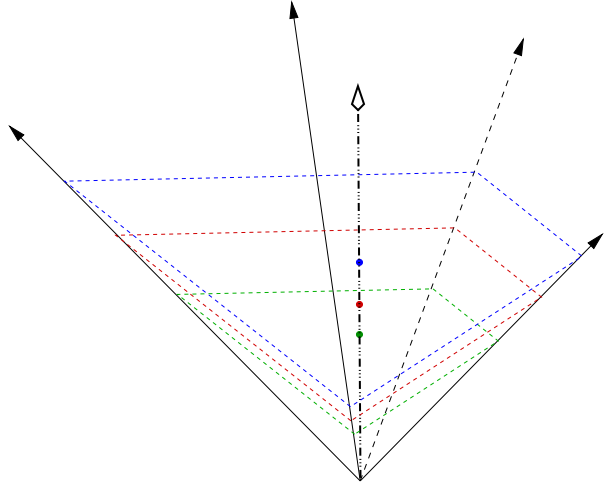}
\caption{A family of parallel planar constraints intersecting the
dual toric cone. All constraint slices are similar polygons, so
their centers of gravity fall on a ray extending from the tip of
the dual toric cone. Here the toric variety is $dP_1$. The
constraint slices are normal to the Reeb vector and represent
equal $R$ surfaces. The diagram was drawn in the standard
$SL(3,\mathbb{Z})$ frame, so that the centers of gravity of equal
R-charge slices fall directly above the tip of the cone.}
\label{maxline}
\end{center}
\end{figure}

In some practical situations, finding the center of gravity of an infinitesimally thin slice $(\vec{s},\vec{n}) = c$ of the dual
toric cone may still be harder than solving another auxiliary problem. Consider a family of parallel slices (constraints) of the
form $(\vec{s},\vec{n}) = c_i$, indexed by $i$. Because $\mathcal{C}^*$ is a cone, all such slices are similar polygons,
differing only by an overall scale. This means that their centers of gravity all fall on the same straight line $k(\vec{s})$,
which is uniquely determined by the normal vector $\vec{s}$ of the constraint plane.  The intersection of this line with the
constraint $(\vec{s},\vec{n}) = c$ then yields the maximal degeneracy point on the plane. This is illustrated in Figure
\ref{maxline}.

Now consider the pyramid $\Delta_{\vec{s},\, c}$ cut out from the dual toric cone by the plane of constraint $(\vec{s},\vec{n}) =
c$. Its center of gravity also evidently lives on the line $k(\vec{s})$. We may now reverse the argument and find the line of
maximal entropy, for this family of constraints given by $(\vec{s},\vec{n}) = c_i$, by finding the center of gravity of the
pyramid $\Delta_{\vec{s},\, c}$. In particular, $k(\vec{s})$ is the ray extending from the tip of the cone in the direction of
the center of gravity of $\Delta_{\vec{s},\, c}$. In this way the point of maximum entropy on the constraint $(\vec{s},\vec{n}) =
c$ may be mechanistically recovered by hanging the pyramid by its tip using a piece of string. When the solid is balanced, the
maximal entropy point will be directly below the tip.

The latter observation leads to a highly peculiar, mechanistic
characterization of the Reeb vector. Consider the dual toric cone
in standard presentation and align the $z$-axis ($x_d$-axis in $d$
dimensions) with the vertical direction. Of all families of
pyramids $\Delta_{\vec{s},\, c}$, the family $\Delta_{\vec{b},\,
c}$ is the unique one whose centers of gravity will lie on the
vertical axis, i.e., directly above the tip. In other words, if we
chop the dual cone $\mathcal{C}^*$ with a plane normal to
$\vec{b}$, the resulting pyramid will balance on its tip!

\subsubsection{Examples} \Label{sec:examples}

\begin{figure}[t]
\begin{center}
\includegraphics[scale=0.4]{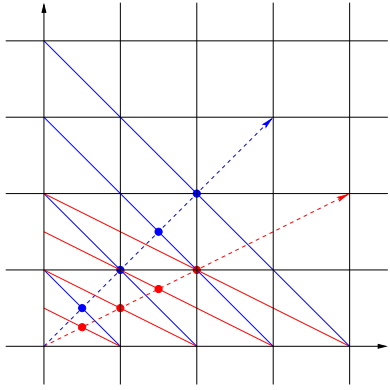}
\caption{Two sets of constraints on the dual toric cone of $\mathbb{C}^2$. Level sets of
the R-charge (normal to $\vec{b}=(1,1)$) and of the quantity $x+2y=\frac{3}{2}R+y$ are given in blue and red,
respectively. The midpoints of each set of line segments (bold dots) trace rays of
fastest entropy growth for each set of constraints (dotted vectors).}\label{fig:c2}
\end{center}
\end{figure}

We begin with the simplest possible example: $\mathbb{C}^2$ and
take the dual toric cone to be the first quadrant of
$\mathbb{Z}^2$, which is not the standard presentation. Here the
Reeb vector is given by $\vec{b}=(1,1)$, and thus the lines
corresponding to the constraint of fixed R charge are normal to
$(1,1)$, i.e., they have slope $-1$.  Here every equal R-charge
constraint is a line segment normal to the Reeb vector
$\vec{b}=(1,1)$. The center of gravity of each line segment is
simply its midpoint, and the union of all midpoints forms the
fastest growth ray $k(\vec{b}) = (1,1)$. This is drawn in blue in
Fig.~\ref{fig:c2}. The numeric equality between $\vec{b}=(1,1)$
and $k(\vec{b})$ is purely accidental\footnote{~These two objects
should not be compared, as $\vec{b}=(1,1)$ lives in the toric cone
while $k(\vec{b})$ lives in the dual toric cone.} and is not
respected by $SL(2,\mathbb{Z})$ transformations. One may wish to
impose a different constraint, for example of the form $x+2y =$
const. The ray of maximal entropy is again given by the union of
the midpoints of the line segments $x+2y=c$ contained in the dual
toric cone, and we can  read off $k((1,2))=(2,1)$. This is
illustrated in red in Figure~\ref{fig:c2}.

As a next example, consider the simplest three-dimensional toric
variety. In its simplest presentation, where both the toric and
the dual toric cone are simply the first octant of the lattice
$\mathbb{Z}^3$, the symmetry relating the three generators is
evident and trivially $k(\vec{b})=(1,1,1)$. For a less trivial
example, consider the presentation of the toric data of
$\mathbb{C}^3$ in the conventions of Section~\ref{sec:review},
where we choose the $SL(3,\mathbb{Z})$ frame so that the
generators of the toric cone reside in the $z=1$ plane. It is
straightforward to compute the generators of the dual toric cone
and the Reeb vector using the results from Section
\ref{sec:review}, and one gets
\begin{equation}
w_1 =  (0,1,0), \quad w_2 = (-1,-1,1), \quad w_3 = (1,0,0), \quad
\vec{b} = (1,1,3).
\end{equation}
Choosing the slicing corresponding to fixed R charge, i.e., the
one corresponding to planes normal to the Reeb vector, we find
that the center of gravity of the pyramid cut out from
$\mathcal{C}^*$ by the plane of R-charge $R$,
$x+y+3z=\frac{3R}{2}$, is the point $(0,0,\frac{3R}{8})$.
Therefore the line of maximal entropy at fixed R-charge is given
by
\begin{equation}
\vec{k}(\vec{b}) = \vec{k}((1,1,3)) = (0,0,1).
\end{equation}
By the argument below eq.~(\ref{reebexplicit}), this holds for
every toric variety in the standard presentation. We remind the
reader that $\vec{k}(\vec{s})$ is to be understood as specifying a
ray, so its normalization is immaterial. The ray intersects the
equal $R$ surface at
\begin{equation}
(0,0,\frac{R}{2}) = \frac{R}{2}(w_1+w_2+w_3)\, ,\Label{maxR2}
\end{equation}
in agreement with eq.~(\ref{maxR}). The form of (\ref{maxR2}),
symmetric in the generators $w_1,\,w_2,\,w_3$, is a direct
consequence of the fact that the three adjoint scalars $X,\,Y,\,Z$
have equal R-charge. As exemplified here, all algorithms presented
in the current section respect the $SL(3,\mathbb{Z})$ symmetry.


Finally, we treat the case of $dP_1$, which was used for
illustration throughout Section~\ref{sec:review}.   Again consider
the family of equal R-charge constraints, which in the (standard)
$SL(3,\mathbb{Z})$ frame of Figure~\ref{fig:toric} take the form
\begin{equation}
(4 - \sqrt{13})y + 3z = \frac{3}{2}R_0\, . \Label{constraintRdP1}
\end{equation}
The center of gravity of the pyramid cut out from $\mathcal{C}^*$
with the plane (\ref{constraintRdP1}) again resides at
$(0,0,\frac{3R_0}{8})$; indeed, this will hold for all pyramids
formed by equal R-charge surfaces in the standard presentation.
The most entropic charge vector for a fixed R-charge $R_0$ takes
the form (\ref{maxR}).

The appearance of an irrational number in the Reeb vector of
$dP_1$, eq.~(\ref{eq:DPReeb}), implies that an \emph{exact}
specification of the R-charge traces a line segment and not a
plane intersecting $\mathcal{C}^*$.   Namely, all the integer
points in the dual toric cone that line within a plane
perpendicular to the Reeb vector  are actually collinear.  In
detail, writing down
\begin{equation}
R_0 = R_1 + R_2\,\sqrt{13}\, , \Label{exactmatch}
\end{equation}
the locus of equal R-charge takes the form
\begin{eqnarray}
4y+3z & =& \frac{3}{2}R_1\nonumber \\
-y & = & \frac{3}{2}R_2\, ,
\end{eqnarray}
which is a line in three-space. However, such considerations would
only be of practical import if one were capable of making
\emph{exact} R-charge (conformal dimension) measurements, so as to
extract from one measurement two integers $R_1,\,R_2$. At any
finite precision $\epsilon$, the relevant constraint to impose
instead is
\begin{equation}
R_0 - \epsilon < R < R_0 + \epsilon\, .\Label{thinshell}
\end{equation}
In this case, the most entropic point is given by the center of
mass of the thin shell (\ref{thinshell}) and is given by
(\ref{maxR}). For completeness, we present in
Appendix~\ref{irrational} an algorithm for calculating the
multiplicity of operators matching a specified irrational R-charge
\emph{exactly} (analogously to eq.~(\ref{meinardus2})).   This
generalizes the construction of Sec.~\ref{partplanar}.

\subsection{Dual giant interpretation}
\llabel{sec:dualmicro}

Recalling the construction \cite{sparks-Z} reviewed in
Sec.~\ref{sec:dualphases}, wherein $\mathcal{C}^*$ is seen as the
phase space of a dual giant graviton propagating in the ${\rm
AdS}_5 \times X$ background, one can also interpret the present
results in terms of dual giant configurations. Dual giant
gravitons are point-like objects from the viewpoint of the
Sasaki-Einstein manifold $X$ and carry momentum $\vec{P} =
(P_{\phi_1},P_{\phi_2},P_{\phi_3})$ along the directions of the
maximal torus of $X$. This momentum is quantized and takes integer
values falling in the dual toric cone $\mathcal{C}^*$. In this
way, the conserved momentum of each dual giant graviton $\vec{P}$
corresponds to a charge vector $\vec{m}$ comprising the three
$U(1)$ (flavor) charges studied above. Likewise, we may think of a
vector $\vec{s} \in \mathcal{C}$ as specifying a particular cycle
$\sigma_{\vec{s}}$ in the maximal torus, and a planar surface
\begin{equation}
(\vec{s}, \vec{P}) = P_{\vec{s}}
\end{equation}
consists of all states whose component of momentum along
$\sigma_{\vec{s}}$ is fixed at $P_{\vec{s}}$. The result of Sec.
\ref{partplanar}, eq.~(\ref{meinardus2}), gives the total number
of such configurations.

Analogously, Sec.~\ref{partind} gives a way of enumerating the
configurations of many dual giants which are characterized by a
given triple of total momenta $\vec{P} =
(P_{\phi_1},P_{\phi_2},P_{\phi_3})$. Following the procedure,
eq.~(\ref{fvol}) assigns to $\vec{P}$ a special cycle
corresponding to $\vec{\nabla} f(\vec{P}) \in \mathcal{C}$.
Eq.~(\ref{meinardus3}) then shows that fixing the total momentum
along that cycle becomes in the asymptotic regime equivalent to
fixing the momentum vector at $\vec{P}$. The reason for that is
reminiscent of the equipartition theorem and may be traced
directly to the central limit theorem. In particular, the
orientation of the cycle $\sigma_{\vec{\nabla} f(\vec{P})}$ in the
maximal torus of $X$ is such that momentum running along it splits
among the fundamental cycles of the torus precisely in the ratio
given by $\vec{P}$.

\section{Universal distribution of trace factors}\setall
\llabel{sec:structure-dist}

In the previous section we found the multiplicity of
operators for any given triple of charges $\vec{n}$, comprising
(linear combinations of) the R-charge and the two flavor charges.
Further, we gave an algorithm which selects the combination of
flavor charges $\vec{n}$ that maximizes this multiplicity given any linear constraint on the charges.  In this section we will show that when $|\vec{n}|$ is large, most of these operators share a universal trace structure.

In \cite{babel}, the structure of multi-trace operators in the
half-BPS sector of $\mathcal{N}=4$ SYM theory was treated using
canonical partition function techniques. The problem of counting
BPS multi-trace operators in an $\mathcal{N}=1$ SCFT living on a
finite stack of D-branes transverse to the tip of a Calabi-Yau
cone was likewise treated in \cite{plethystic} using the
plethystic program. In the following we shall proceed along
similar lines in order to extract statistical information about
the distribution of traces in a multi-trace operator with charge
vector $\vec{n}$.

As explained in Section~\ref{sec:entropy}, every multi-trace
operator of charge vector $\vec{n}$ is specified by a vector
partition
\begin{equation}
\sum_j \vec{m}_{j} = \vec{n}\, ,
\end{equation}
where each summand $\vec{m}_j \in \mathcal{C}^*$ corresponds to one trace factor, unique up to F-term relations.  The exact
correspondence between flavor charges, winding paths in the dimer model, and fields entering the given trace factor, was
described in Section~\ref{sec:review}. Introducing a triple of generalized temperatures\footnote{~These temperatures are not
physical. They are Lagrange multipliers introduced for ease of calculation.} $\vec{\beta}\equiv (\beta_1,\beta_2,\beta_3)$ and a
chemical potential $\alpha$ , we may write down a grand canonical partition function for the system \cite{plethystic,sparks-Z}
\begin{equation}
Z = \prod_{\vec{m} \in \mathcal{C}^*} \big(1 -
e^{\alpha-(\vec{\beta},\, \vec{m})}\big)^{-1} \equiv
\prod_{\vec{m} \in \mathcal{C}^*} Z_{\vec{m}} \, .
\Label{partitiongen}
\end{equation}
This partition function has been exhibited in \cite{sparks-Z} as
counting dual giant graviton states moving on a toric variety
$\mathcal{M}$, and concurrently in \cite{plethystic} as a
generating function for mesonic multi-trace operators in the
associated field theory. In the following, we first treat the case
where $\alpha =0$ so that every summand $\vec{m}$ (trace factor of
type $\vec{m}$) is counted with the same weight. Indeed, each
factor $Z_{\vec{m}} = \big(1 - e^{-(\vec{\beta},\,
\vec{m})}\big)^{-1}$ allows for an arbitrary number of repetitions
of the summand $\vec{m}$ in a partition (multi-trace operator).
Thus, in setting $\alpha=0$, we are going to let the ensemble
choose its optimal number of trace factors $M$. The case where $M$
is fixed by hand with a non-trivial chemical potential is treated
later in the section.

\subsection{Ensemble with unrestricted number of traces}
\llabel{sec:mfree}

With $\alpha = 0$, the partition function (\ref{partitiongen})
blows up at $\vec{0} \in \mathcal{C}^*$. This corresponds to the
onset of Bose-Einstein condensation at the trivial point. However,
the tip of the cone corresponds to a trivial factor in the
structure of a multi-trace operator, and as such should not be
counted in the total number of factors in a multi-trace operator.
Therefore, the correct partition function to be analyzed is:
\begin{equation}
Z = \prod_{\vec{m} \in \mathcal{C}^* \backslash \vec{0}} \big(1 -
e^{-(\vec{\beta},\, \vec{m})}\big)^{-1} \equiv \prod_{\vec{m} \in
\mathcal{C}^* \backslash \vec{0}} Z_{\vec{m}} \, .
\Label{partition1}
\end{equation}
For a given charge vector $\vec{n}$, the Lagrange multipliers
$\vec{\beta}$ are set by requiring:
\begin{equation}
\langle n_i \rangle = -\frac{\partial \log{Z}}{\partial \beta_i}\quad\quad i =
1,2,3\, .\Label{setbeta}
\end{equation}
Because eq.~(\ref{partition1}) factorizes, we can read off the
mean and standard deviation in the population of the trace factor
$\vec{m}$:
\begin{equation}
\langle n_{\vec{m}} \rangle = -\frac{1}{m_i}\frac{\partial
\log{Z_{\vec{m}}}}{\partial \beta_i} =
(e^{(\vec{\beta},\,\vec{m})}-1)^{-1}\,. \Label{meanocc}
\end{equation}
The right hand side is manifestly independent of $i$. Indeed, (\ref{setbeta}) counts average charge while (\ref{meanocc}) counts the average number of traces;
their relation is
\begin{equation}\label{n-nm}
\langle n_i \rangle = \sum_{\vec{m} \in \mathcal{C}^* \backslash \vec{0}}
m_i \langle n_{\vec{m}} \rangle \ .
\end{equation}
The standard deviation in $n_{\vec{m}}$ reads:
\begin{equation}
\sigma(n_{\vec{m}}) = \Big(\langle n_{\vec{m}}^2 \rangle - \langle
n_{\vec{m}} \rangle^2\Big)^{1/2} = \Big(\frac{1}{m_i^2
Z_{\vec{m}}}\frac{\partial^2 Z_{\vec{m}}}{\partial \beta_i^2} -
\big(e^{(\vec{\beta},\,\vec{m})}-1\big)^{-2}\Big)^{1/2} =
\Big(2\sinh{\frac{(\vec{\beta},\,\vec{m})}{2}}\Big)^{-1}\,
,\Label{stdevnm}
\end{equation}
so that the standard deviation always exceeds the mean, most
rampantly away from the tip of the cone:
\begin{equation}
\sigma(n_{\vec{m}}) / \langle n_{\vec{m}} \rangle =
e^{\frac{(\vec{\beta},\,\vec{m})}{2}} = \big(1 + \langle
n_{\vec{m}} \rangle^{-1} \big)^{1/2} \geq 1\, .\Label{ratiostdev}
\end{equation}
This is the same relation as that found in \cite{babel}. We defer
a full discussion of the structure of heavy multi-trace operators
until the next subsection.

\begin{figure}[t]
\begin{center}
\includegraphics[scale=0.4]{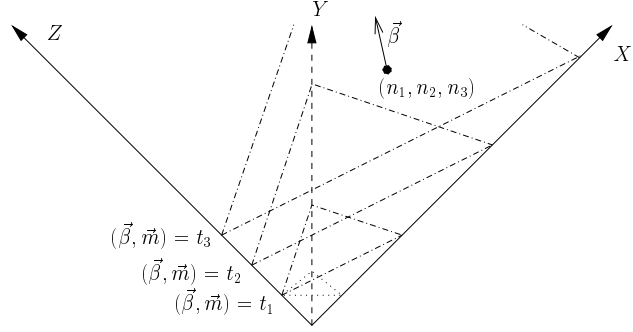}
\caption{Illustration of the natural grading associated to the
point $(n_1,n_2,n_3)$ for $\mathbb{C}^3$. The planes with dashed
lines are normal to $\vec{\beta}$, and for comparison we have
included the level 1 surface containing Tr $X$, Tr $Y$ and Tr $Z$
with dotted lines. } \label{fig:beta}
\end{center}
\end{figure}

Eq.~(\ref{meanocc}) shows that the planar surfaces
\begin{equation}
(\vec{\beta},\vec{m}) = y \Label{eqplane}
\end{equation}
share the same occupation numbers. This induces a natural grading
of the dual cone $\mathcal{C}^*$ by parallel planes with the
normal $\vec{\beta}$; we illustrate this in  Figure
\ref{fig:beta}. It is easy to see that these planes are the same
as the planes $p_m$ utilized in Sections~\ref{partplanar},
\ref{partind} to evaluate the number of vector partitions of a
given vector $\vec{n}$.\footnote{~This follows from the
observation that the counting procedure in
Sections~\ref{partplanar}, \ref{partind} did not distinguish among
charge vectors lying on the same plane $p_m$. Thus, the planes
$p_m$ comprise vectors of equal multiplicity, as do the planes
(\ref{eqplane}), so the two families must coincide.} Consequently
$\vec{\beta}
\parallel \vec{\nabla}f(\vec{n})$ and we may recast the
populations at sites $\vec{m}$ in terms of just one temperature
$\beta$:
\begin{equation}
\langle n_{\vec{m}} \rangle =
-\frac{1}{(\vec{m},\,\vec{\nabla}f(\vec{n}))}\frac{\partial
\log{Z_{\vec{m}}}}{\partial \beta} =
(e^{\,\beta\,(\vec{m},\,\vec{\nabla}f(\vec{n}))}-1)^{-1}\,,
\Label{meanocc2}
\end{equation}
where we have set
\begin{equation}
\vec{\beta} \equiv \beta\,\vec{\nabla}f(\vec{n})\, .
\Label{defbeta}
\end{equation}
With this form of $\vec{\beta}$ substituted in
eq.~(\ref{partition1}), the condition setting $\beta$ reads:
\begin{equation}
(\vec{n},\,\vec{\nabla}f(\vec{n})) = -\frac{\partial
\log{Z}}{\partial \beta}\, . \Label{setbeta1}
\end{equation}

For large $|\vec{n}|$, we can use a continuum approximation. A
thin shell contained between two planes normal to
$\vec{\nabla}f(\vec{n})$ given by
\begin{equation}
t < (\vec{m},\,\vec{\nabla}f(\vec{n})) < t + dt \Label{deft}
\end{equation}
is seen from eq.~(\ref{eq:am}) to contain
\begin{equation}
a_t\, dt =  \frac{V_t\, d}{t} dt
\end{equation}
lattice points. In order to write an explicit expression for $V_t$
in $d$ dimensions, we shall need further notation. Write the dual
toric cone $\mathcal{C}^*$ as a union of $F$ simplicial cones
$\hat{\sigma}^i$ and denote the $d$ generators of $\hat{\sigma}^i$
by $\hat{\vec{w}}^{\,i}_j$, $j=1,\,\ldots,\,d$. Each of these is a
$d$-vector, so $(\hat{w}^i)_{jk}$ is a $d\times d$ matrix. In
terms of the notation of Sec.~\ref{sec:toricfan}, which is
suitable to $d=3$, one choice of the $F$ simplicial cones leads
to:
\begin{eqnarray}
\hat{\vec{w}}^{\,i}_1 & = & \vec{w}_{k-1,0} \nonumber \\
\hat{\vec{w}}^{\,i}_2 & = & \vec{w}_{i-1,i} \\
\hat{\vec{w}}^{\,i}_3 & = & \vec{w}_{i,i+1} \nonumber
\end{eqnarray}
Here $i$ ranges from 0 to $F= k-2$, where the Newton polygon is a
$k$-gon. The volume of the dual toric cone bound by the plane
$(\vec{s},\vec{x})=t$ with normal $\vec{s}$ is then given by
\begin{equation}
V_t = \frac{t^d}{d!} \,\sum_{i=1}^F
\frac{|\det{\hat{w}^i|}}{\prod_{j=1}^d
(\hat{\vec{w}}^{\,i}_j,\vec{s})}\, . \Label{volume}
\end{equation}
It follows that the thin shell (\ref{deft}) contains
\begin{equation}
\frac{1}{(d-1)!}\sum_{i=1}^F
\frac{|\det{\hat{w}^i|}}{\prod_{j=1}^d
(\hat{\vec{w}}^{\,i}_j,\vec{\nabla}f(\vec{n}))}\, t^{d-1} dt
\equiv B(\vec{n})\, t^{d-1}dt \Label{at}
\end{equation}
lattice points. Recalling from the definition (\ref{fvol}) that
$f(\vec{n}) = V_t\,d$ for $t =
(\vec{n},\,\vec{\nabla}f(\vec{n}))$, we see that $B(\vec{n})$ may
be related to $f(\vec{n})$ as
\begin{equation}
B(\vec{n}) = \frac{V_t \,d}{t^d} =
\frac{f(\vec{n})}{(\vec{n},\,\vec{\nabla}f(\vec{n}))^d} \ ,
\end{equation}
that is,
\begin{equation}
f(\vec{n})
= (\vec{n},\vec{\nabla}f(\vec{n}))^d\, B(\vec{n})
= \frac{(\vec{n},\vec{\nabla}f(\vec{n}))^d}{(d-1)!}
\,\sum_{i=1}^F \frac{|\det{\hat{w}_i|}}{\prod_{j=1}^d
(\hat{\vec{w}}_{ij},\,\vec{\nabla}f(\vec{n}))} \ .
\Label{auxfb}
\end{equation}

Now, each of the points in the thin shell has mean population
$(\exp{\beta\,t}-1)^{-1}$, and contributes $t$ to the right hand
side of eq.~(\ref{setbeta1}), as a continuous version of
(\ref{n-nm}). Therefore, we may approximate the condition
(\ref{setbeta1}) for $\beta$ by an integral:
\begin{equation}
B(\vec{n}) \int_0^{\infty}
\frac{t^d}{e^{\beta\, t}-1} \, dt =
(\vec{n},\,\vec{\nabla}f(\vec{n}))\, , \Label{setbeta2}
\end{equation}
where the negligible contribution for large $t$ was added to the
integral to facilitate explicit evaluation. Evaluating the
integral, we get
\begin{equation}
\beta^{d+1} = \frac{B(\vec{n}) \, \Gamma(d+1)\, \zeta(d+1)}
{(\vec{n},\,\vec{\nabla}f(\vec{n}))}\, , \Label{setbeta3}
\end{equation}
which, using eqs.~(\ref{meinardus3}) and (\ref{auxfb}), yields an identity:
\begin{equation}
\beta =
\frac{d}{d+1}\,\frac{S(\vec{n})}{(\vec{n},\,\vec{\nabla}f(\vec{n}))}
\Label{setbeta3half}
\end{equation}
Eq.~(\ref{setbeta3}) is in agreement with the well known case
$d=1$.  There $f(n)=n$ so that $B(n)=1$ and $\beta^2=\zeta(2)/n$,
as in eq. (3.7) of \cite{thermal}.  In the case of primary
interest, $d=3$, in the notation of Sec.~\ref{sec:toricfan},
$\beta$ takes the form:
\begin{equation}
\beta = \pi\, \Big(
\frac{1}{30\,(\vec{n},\,\vec{\nabla}f(\vec{n}))\,(\vec{\nabla}f(\vec{n}),\vec{w}_{k-1,0})}\,
\sum_{i=1}^{k-2}
\frac{(\vec{w}_{k-1,0},\vec{w}_{i-1,i},\vec{w}_{i,i+1})}
{(\vec{\nabla}f(\vec{n}),\vec{w}_{i-1,i})\,(\vec{\nabla}f(\vec{n}),\vec{w}_{i,i+1})}
\Big)^{1/4} \Label{setbeta4}
\end{equation}

\subsubsection{Structure of heavy multi-trace operators}

Eqs.~(\ref{meanocc2}, \ref{at}, \ref{setbeta3}) allow us to
uncover a wealth of information about the structure of a typical
large charge multi-trace operator. We list such observations
below.

The average number of trace factors in the operator is given by:
\begin{equation}
\langle M (\vec{n})\rangle_0 = B(\vec{n}) \int_0^{\infty}
\frac{t^{d-1}}{e^{\beta\, t}-1} \, dt =
\frac{B(\vec{n})\,\Gamma(d)\,\zeta(d)}{\beta^d} \Label{avtrace}
\end{equation}
Now, using the relation between $f(\vec{n})$ and $B(\vec{n})$ from (\ref{auxfb}), we therefore have
\begin{eqnarray}
\langle M (\vec{n}) \rangle_0 & = &
\frac{B(\vec{n})\,\Gamma(d)\,\zeta(d)}{\left(
B(\vec{n})\,\Gamma(d+1)\,\zeta(d+1) \right)^{\frac{d}{d+1}}}
(\vec{n},\vec{\nabla}f(\vec{n}))^{\frac{d}{d+1}} \nonumber \\
& = & \frac{B(\vec{n})\,\Gamma(d)\,\zeta(d)}{\left(
B(\vec{n})\,\Gamma(d+1)\,\zeta(d+1) \right)^{\frac{d}{d+1}}}
\left( \frac{f(\vec{n})}{B(\vec{n})} \right)^{\frac{1}{d+1}}
\nonumber \\
& = & \frac{\zeta(d)}{d\,\zeta(d+1)} \Big(f(\vec{n})\, \Gamma(d +
1)\, \zeta(d+1)\Big)^{\frac{1}{d+1}} =
\frac{\zeta(d)}{(d+1)\,\zeta(d+1)} \,S(\vec{n})\,.
\Label{proportionms}
\end{eqnarray}
In the first line we substituted for $\beta^{-d}$ from
eq.~(\ref{setbeta3}) and in the second for
$(\vec{n},\vec{\nabla}f(\vec{n}))$ from eq.~(\ref{auxfb}). The
average number of traces is directly proportional to the entropy,
with the constant of proportionality determined solely by the
dimension of the toric variety $d$.

\begin{figure}[t]
\begin{center}
\includegraphics[scale=0.6]{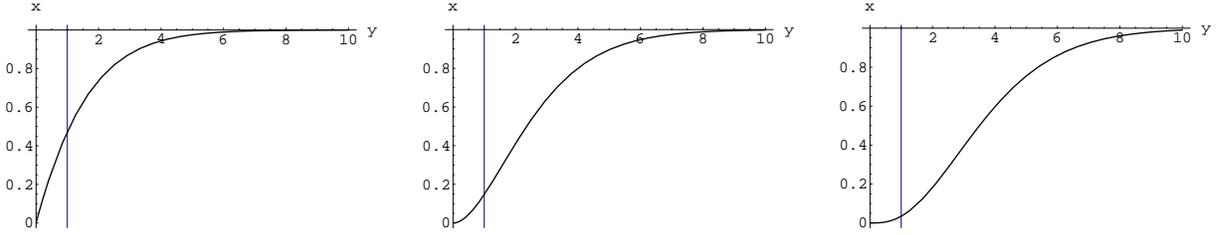}
\caption{Left to right: $x_2(y)$, $x_3(y)$, $x_4(y)$.}
\label{fig:tableaux}
\end{center}
\end{figure}

The mean multiplicity $\langle n_{\vec{m}} \rangle$
of the $\vec{m}$-trace in the large
multi-trace operator, eq.~(\ref{meanocc2}), quickly drops to 0 as
$|\vec{m}|$ increases. The only traces whose occupations are of
the order of 1 or greater are those satisfying
\begin{equation}
\beta\, (\vec{m},\,\vec{\nabla}f(\vec{n})) \,\lesssim\,
\mathcal{O}(1)\, . \Label{cond1}
\end{equation}
By eq.~(\ref{ratiostdev}), the same relation with a $\ll$ sign
defines the regime where the mean \emph{accurately} predicts a
non-zero occupancy. Substituting for $\beta$ from
eqs.~(\ref{setbeta3}, \ref{setbeta3half}), we have:
\begin{equation}
\frac{(\vec{m},\,\vec{\nabla}f(\vec{n}))}{(\vec{n},\,\vec{\nabla}f(\vec{n}))}\,
\lesssim\, \frac{d+1}{d\,S(\vec{n})} \,
\Longleftrightarrow\, (\vec{m},\,\vec{\nabla}f(\vec{n})) \lesssim
\Big(
\frac{(\vec{n},\,\vec{\nabla}f(\vec{n}))}{\Gamma(d+1)\,\zeta(d+1)\,B(\vec{n})} \Big)^{\frac{1}{d+1}}\, .
\Label{mregime}
\end{equation}
This scaling behavior motivates the introduction of an auxiliary
variable
\begin{equation}
y(\vec{m}) \equiv \beta\, (\vec{m},\,\vec{\nabla}f(\vec{n}))\, .
\end{equation}
In comparison with the variable $t$ of eq.~(\ref{deft}), $y$ is
rescaled by $y = \beta t$. It is clear that mean occupation
numbers and all other quantities which are extractable from
$Z(\beta)$ are functions of $y$ alone. It is convenient to
introduce the quantity
\begin{equation}
x(y) = \int_0^y \frac{\tilde{y}^{d-1}
d\tilde{y}}{e^{\tilde{y}}-1}\, \Bigg/ \, \int_0^\infty
\frac{\tilde{y}^{d-1} d\tilde{y}}{e^{\tilde{y}}-1}\,. \Label{defx}
\end{equation}
By the same logic which led to eqs.~(\ref{setbeta2},
\ref{avtrace}), the quantity $x(y)$ gives that fraction of trace
factors making up a heavy multi-trace operator (given by
(\ref{avtrace})) which have $\tilde{y} < y$. The graph of $x(y)$
is therefore a convenient presentation of \emph{all the universal
features} of a heavy multi-trace mesonic BPS operator.
Importantly, $x(y)$ depends only on the dimension of the toric
variety $d$, which motivates the notation $x_d(y)$. This leads to
the interesting conclusion that in all world-volume SCFTs on
D3-branes transverse to tips of toric Calabi-Yau threefolds, all
heavy BPS operators are described by the same curve $x_3(y)$. The
curves $x_2(y)$, $x_3(y)$, and $x_4(y)$ are plotted in Figure
\ref{fig:tableaux}.

The graphs $x_d(y)$ are interpreted as follows (we leave a speculative
interpretation to Appendix \ref{ap:amoeba}).
Suppose one is interested in the structure of a
heavy multi-trace mesonic BPS operator with charge vector
$\vec{n}$. One begins by computing $B(\vec{n})$ from
eq.~(\ref{at}) and $\beta$ from eq.~(\ref{setbeta3}). The number
of trace factors in equivalence classes $\vec{m}$ satisfying
\begin{equation}
\beta^{-1}y \,<\, (\vec{m},\,\vec{\nabla}f(\vec{n})) \,<\,
\beta^{-1}(y+dy)
\end{equation}
is given by
\begin{equation}
\frac{B(\vec{n})}{\beta^{d}}\, \int_0^\infty \frac{\tilde{y}^{d-1}
d\tilde{y}}{e^{\tilde{y}}-1}\, \left( \frac{dx}{dy} \right) dy\, .
\end{equation}
Here we have differentiated eq.~(\ref{defx}) and used
eq.~(\ref{avtrace}) with $\tilde{y} = \beta t$. Using
eq.~(\ref{at}) one may also evaluate the average number of trace
factors corresponding to an individual site $\vec{m}$. This is
given by
\begin{equation}
\langle n_{\vec{m}} \rangle = \langle n_{y(\vec{m})} \rangle =
\frac{1}{y(\vec{m})^{d-1}} \left( \frac{dx}{dy}
\right) \int_0^\infty \frac{\tilde{y}^{d-1}
d\tilde{y}}{e^{\tilde{y}}-1}\,. \Label{meanocc3}
\end{equation}
Curiously, mean occupations per site depend only on the
orientation of the vector $\vec{n}$ in $\mathcal{C}^*$.

Since the quantity $x_d(y)$ stands for the fraction of trace
factors with $\tilde{y}<y$, it varies between 0 and 1. In
particular, $1-x(y)$ is the fraction of operators with $\tilde{y}$
exceeding $y$. From the fact that $x(y)$ approaches 1 very rapidly
for $y$ of the order of unity one sees that \emph{almost all}
summands satisfy $|\vec{m}| \propto |\vec{n}|^{\frac{1}{d+1}}$.
The characteristic scale on which the occupations (\ref{meanocc3})
vary is thereby set by the quantity $|\vec{n}|^{\frac{1}{d+1}}$.

Recalling eq.~(\ref{cond1}), the sites with occupations of the
order of 1 or greater correspond to the section of the graph with
$y<1$. These are also the sites for which the mean occupancies
(\ref{meanocc2}, \ref{meanocc3}) are good predictors for actual
occupation numbers. Since $x_3(1) \approx 0.15$, we see that
approximately 85\% of the trace factors entering a heavy operator
are randomly selected from a large volume in $\mathcal{C}^*$. The
size of that volume is such that $y>1$ but still of the order of
unity. In particular, $x_3(10) \approx 0.998$, which means that on
average only 1 in 500 trace factors has
$(\vec{m},\,\vec{\nabla}f(\vec{n})) >
10\,(\vec{n},\,\vec{\nabla}f(\vec{n}))^{\frac{1}{d+1}}$.

Denote the inverse of $x_d(y)$ as $y_d(x)$, so that
$y_d(x_d(\tilde{y})) = \tilde{y}$. Consider the differential
segment $(x,\, x+dx)$. It stands for $\langle M(\vec{n})
\rangle_0\, dx$ traces, each contributing $\beta^{-1}y_d(x)$ to
the right hand side of eq.~(\ref{setbeta2}). Therefore, $y_d(x)\,
dx$ is proportional to that fraction of the total charge of the
heavy operator, which is carried by the traces with $y_d(x) <
\tilde{y} < y_d(x+dx)$. In particular, the charge carried by the
trace operators with $\tilde{y}$ not exceeding $y$ is given by
\begin{equation}
R'(\tilde{y}<y) \propto \int_0^{x_d(y)} \tilde{y}_d(x)\,dx \propto \int_0^y
\tilde{y} \frac{\tilde{y}^{d-1}}{e^{\tilde{y}}-1}\,d\tilde{y}
\,\propto\, x_{d+1}(y)\,. \Label{rproportion}
\end{equation}
Here we performed a simple change of variable using the definition
of $x_d(y)$, eq.~(\ref{defx}). Eq.~(\ref{rproportion}) holds for
any charge of interest $R'$ because for any linear combination of
the $d$ $U(1)$ charges the total charge grows proportionately as
one includes factors of increasing $y$. In conclusion, the
function $x_{d+1}(y)$ gives the fraction of the total charge of
the operator residing in trace factors with $\tilde{y} < y$. As an
example, $x_4(1) \approx 3\%$, which means that only 3\% of the
total charge of a heavy operator is carried by trace factors whose
expected occupancies are of the order unity or greater. The
remainder is contained in trace factors which are randomly pooled
from the volume
$(\vec{n},\,\vec{\nabla}f(\vec{n}))^{\frac{1}{d+1}} < \beta^{-1}y
< \kappa \,(\vec{m},\,\vec{\nabla}f(\vec{n}))^{\frac{1}{d+1}}$,
where $\kappa \gtrsim 1$.

\subsection{Ensemble with a fixed number of traces}
\llabel{sec:mnon0} We now turn to the case where $\alpha < 0$. The
condition (\ref{setbeta}) is supplemented by a condition fixing
the number of traces $M$. In keeping with the standard language of
the grand canonical ensemble, we shall alternately refer to trace
factors as the \emph{particles} of the system. The condition sets
the parameter $\alpha$:
\begin{equation}
M = \frac{\partial \log{Z}}{\partial \alpha} \, . \Label{setalpha}
\end{equation}
Here $Z$ is again as given by equation (\ref{partitiongen}) and
includes $\vec{0} \in \mathcal{C}^*$. The results
(\ref{meanocc}-\ref{ratiostdev}) are altered simply by
substituting:
\begin{equation}
(\vec{\beta},\,\vec{m}) \rightarrow (\vec{\beta},\,\vec{m}) -
\alpha \Label{generalize}
\end{equation}
and as a consequence, the planar surfaces (\ref{eqplane}) continue
to comprise points of equal occupations. One consequence of that
is that the validity of Sec.~\ref{centerofgravity} will also
extend to the present section: on each plane $p$ intersecting
$\mathcal{C}^*$, the charge vector of maximal entropy \emph{in the
$M$-trace sector} is the center of gravity of $\{p \cap
\mathcal{C}^* \}$. Furthermore, the argument which motivated
(\ref{defbeta}) also continues to apply. Thus, we shall keep on
referring to the scalar quantity $\beta$ under the identification
$\vec{\beta} \equiv \beta\,\vec{\nabla}f(\vec{n})$.

Proceeding as in Sec.~\ref{sec:mfree}, we approximate
eqs.~(\ref{setbeta}, \ref{setalpha}) by integrals:
\begin{eqnarray}
(\vec{n},\,\vec{\nabla}f(\vec{n})) & = & B(\vec{n})
\int_0^{\infty} \frac{t^d dt}{e^{\beta\, t-\alpha}-1} =
\frac{B(\vec{n})\,\Gamma(d+1)\,g_{d+1}(e^\alpha)}{\beta^{d+1}}
\Label{setbetagen} \\
M & = & B(\vec{n}) \int_0^{\infty} \frac{t^{d-1} dt}{e^{\beta\,
t-\alpha}-1} =
\frac{B(\vec{n})\,\Gamma(d)\,g_d(e^\alpha)}{\beta^d}
\Label{setalphagen}\,,
\end{eqnarray}
where $g_{\nu}(e^{\alpha})$ stand for Bose-Einstein, or
polylogarithm functions. We recall the definition to be
$g_{\nu}(z) := \sum\limits_{k=1}^\infty \frac{z^k}{k^n}$.

Using eq.~(\ref{auxfb}) to eliminate
$\beta$ from the both equations, we get:
\begin{equation}
M = \frac{B(\vec{n})\,\Gamma(d)\,g_d(e^{\alpha})}
{\big(B(\vec{n})\,\Gamma(d+1)\,g_{d+1}(e^{\alpha})\big)^{\frac{d}{d+1}}}
(\vec{n},\,\vec{\nabla}f(\vec{n}))^{\frac{d}{d+1}} =
\frac{\Gamma(d)\,g_d(e^{\alpha})}{\big(\Gamma(d+1)\,g_{d+1}(e^{\alpha})\big)^{\frac{d}{d+1}}}\,
f(\vec{n})^{\frac{1}{d+1}}\, .\Label{setalphaindep}
\end{equation}
Eq.~(\ref{setalphaindep}) allows one to set $\alpha$ independently
of $\beta$. However, the coefficient of
$f(\vec{n})^{\frac{1}{d+1}}$ in (\ref{setalphaindep}) is bounded
from above by
\begin{equation}
\kappa(d) =
\frac{\Gamma(d)\,\zeta(d)}{\big(\Gamma(d+1)\,\zeta(d+1)\big)^{\frac{d}{d+1}}}
\Label{defkappa}
\end{equation}
since $0 \leq e^{\alpha} \leq 1$. For the case of most interest,
$d=3$, we have
\begin{equation}
\kappa(3) \approx 0.59 \, .\Label{kappa3}
\end{equation}
This quantity marks the critical number of trace factors for the
onset of Bose-Einstein condensation. Accordingly, we define
\begin{equation}
M_{\rm crit}(\vec{n}) = \kappa(d)\, f(\vec{n})^{\frac{1}{d+1}}
\Label{mcrit}
\end{equation}
as the critical number of traces for a given charge vector
$\vec{n}$. Note that in the $\alpha = 0$ case, wherein we let the
ensemble choose its number of traces, we had
\begin{equation}
\langle M(\vec{n}) \rangle_0 = M_{\rm crit}(\vec{n}) \, ,
\end{equation}
as can be easily verified from eq.~(\ref{proportionms}). For $M
\neq M_{\rm crit}$, we distinguish the following cases:

\subsubsection{The classical phase}
For $M < M_{\rm crit}(\vec{n})$, $\langle n_{\vec{0}} \rangle$ is
a finite number, so including or excluding the tip of the cone
(contrast the products in eq.~(\ref{partitiongen}) and
eq.~(\ref{partition1})) is immaterial. In this case the condition
(\ref{setalpha}) may equally well be taken to specify the
\emph{exact} number of traces or the \emph{maximal} number of
traces. We call this regime the classical phase, to distinguish it
from the Bose-Einstein phase below. For completeness, we quote the
usual formula for the entropy in the classical phase:
\begin{equation}
S(\vec{n},M) = \log{Z} + \beta (\vec{n},\,\vec{\nabla}f(\vec{n}))
- \alpha M\, .
\end{equation}

\subsubsection{Bose-Einstein phase with trivial
point included} \llabel{sec:incl}

When $M > M_{\rm crit}(\vec{n})$, a macroscopic proportion of
trace factors condense on a small set of points and the system
enters the Bose-Einstein phase. Depending on whether one works
with partition function (\ref{partitiongen}) or
(\ref{partition1}), that set of points is either the tip of the
cone or the points nearest to it.

We begin by including the trivial point and consider partition
function (\ref{partitiongen}). This corresponds to fixing the
\emph{maximal} number of traces at $M$. The requisite counting is
given by the generating function $g_M(\vec{t})$ in
\cite{plethystic}. In this case, the condensate at $\vec{0}$
corresponds to $M - M_{\rm crit}(\vec{n}) > 0$ excess trace
factors, which the system dumps in the immaterial ground state.
The physical meaning of the Bose-Einstein condensate is clear: it
is the phase where it is more entropically favorable for the
system to decrease the number of (non-trivial) trace factors.

Following the standard treatment of Bose-Einstein condensates, we
have
\begin{eqnarray}
(\vec{n},\,\vec{\nabla}f(\vec{n})) & = &
\frac{B(\vec{n})\,\Gamma(d+1)\,\zeta(d+1)}{\beta^{d+1}}\, \\
\alpha &=& -\frac{1}{M_{\rm bc}} + \mathcal{O}(M_{\rm bc}^{-2})\,
,
\end{eqnarray}
where in the second line we assumed $M_{\rm bc}$, the population
of the condensate, to be large. The entropy takes the form:
\begin{eqnarray}
S(\vec{n},M) & = & \log{Z} + \beta
(\vec{n},\,\vec{\nabla}f(\vec{n})) - \alpha (M-M_{\rm bc}) \nonumber \\
& = & \log{Z} + \beta (\vec{n},\,\vec{\nabla}f(\vec{n})) +
\frac{M_{\rm crit}(\vec{n})}{M_{\rm bc}} + \mathcal{O}(M_{\rm
bc}^{-2})\, .\Label{entropywithtrivial}
\end{eqnarray}
In the large $M_{\rm bc}$ limit, only the first two terms remain.
These, of course, correspond to the entropy contained in the
$M_{\rm crit}(\vec{n})$ excited particles (non-trivial traces.)
Therefore, (\ref{entropywithtrivial}) is expected to agree with
(\ref{meinardus3}) by the standard arguments establishing the
equivalence of the microcanonical and grand canonical analyzes in
the thermodynamic limit.

\subsubsection{Bose-Einstein phase with trivial
point excluded} \llabel{sec:excl}

We now work with the partition function
\begin{equation}
Z = \prod_{\vec{m} \in \mathcal{C}^* \backslash \vec{0}} \big(1 -
e^{\alpha-(\vec{\beta},\, \vec{m})}\big)^{-1} \equiv
\prod_{\vec{m} \in \mathcal{C}^* \backslash \vec{0}} Z_{\vec{m}}
\, .
\end{equation}
This case corresponds to fixing the \emph{exact} number of traces
at $M > M_{\rm crit}(\vec{n})$.

Each state in the system is characterized by a choice of
$n_{\vec{m}}$ such that
\begin{equation}
\sum_{\vec{m}} n_{\vec{m}}\, \vec{m} = \vec{n}\, .
\Label{strstate}
\end{equation}
As seen from eqs.~(\ref{meanocc2}, \ref{generalize}), the
occupations $n_{\vec{m}}$ are controlled by the quantity
$y(\vec{m}) = (\vec{m},\,\vec{\nabla}f(\vec{n}))$, which plays the
role of the energy of the system. With the tip of the cone
excluded, the Bose-Einstein condensate will accumulate on those
integral points in $\mathcal{C}^*$ for which $y(\vec{m})$ is the
smallest. We denote this minimal value with $y_0$. The structure
of a state as a vector partition (\ref{strstate}) descends to a
partition of the quantity $y(\vec{n}) =
(\vec{n},\,\vec{\nabla}f(\vec{n}))$:
\begin{equation}
y(\vec{n}) = \sum_{\vec{m}} n_{\vec{m}}\, y(\vec{m}) = M_{\rm
bc}\, y_0 + \sum_{\vec{m} \not\in {\rm bc}} n_{\vec{m}}\,
y(\vec{m}) \, . \Label{strenergy}
\end{equation}
In the last equality we have grouped the contributions to
$y(\vec{n})$ of the Bose-Einstein condensate and of the excited
particles, respectively.

Let $0 < c < 1$ be the proportion of $y(\vec{n})$ contributed by
the second term in (\ref{strenergy}):
\begin{eqnarray}
\sum_{\vec{m} \not\in {\rm bc}} n_{\vec{m}}\, y(\vec{m}) & = & c\,
y(\vec{n}) \qquad\qquad \Rightarrow \qquad\qquad \sum_{\vec{m}
\not\in {\rm bc}} n_{\vec{m}}\, \vec{m} =  c\, \vec{n}
\Label{excitedpart} \\
M_{\rm bc}\, y_0 &=& (1-c) \,y(\vec{n})
\end{eqnarray}
The condensate does not begin to set in until the population of
all the excited particles reaches the critical value. The latter
is given by
\begin{equation}
M_{\rm crit}(c\vec{n}) = c^{\frac{d}{d+1}} M_{\rm crit}(\vec{n})
\end{equation}
since the charge vectors of the excited particles form a partition
of $c\vec{n}$ as in eq.~(\ref{excitedpart}). The scaling of
$M_{\rm crit}(c\vec{n})$ with $c$ follows from the definition
(\ref{mcrit}) and the fact that $f(\vec{n})$ is a $d$-dimensional
volume. Then imposing the condition that the total number of
traces be $M$ is equivalent to requiring:
\begin{equation}
M = (1-c)\frac{y(\vec{n})}{y_0} + c^{\frac{d}{d+1}}M_{\rm
crit}(\vec{n})\, . \Label{setc}
\end{equation}
Using the fact that eq.~(\ref{excitedpart}) is a partition of
$c\vec{n}$ with a critical number of traces, one may feed the
solution of (\ref{setc}) into eq.~(\ref{setbeta2}) to get:
\begin{equation}
c\beta^{d+1} =
\frac{B(\vec{n})\,\Gamma(d+1)\,\zeta(d+1)}{(\vec{n},\,\vec{\nabla}f(\vec{n}))}\,
. \Label{cbeta}
\end{equation}
As before, the chemical potential in the large $M_{\rm bc}$ approximation
is given by
\begin{equation}
\alpha \approx -\frac{1}{M_{\rm bc}} =
-(1-c)^{-1}\frac{y_0}{y(\vec{n})}\, .
\end{equation}
This yields for the entropy
\begin{eqnarray}
S(\vec{n},M) & = & \log{Z} + c\,\beta
(\vec{n},\,\vec{\nabla}f(\vec{n})) - \alpha (M-M_{\rm bc}) \nonumber \\
& \approx & \log{Z} +
\frac{B(\vec{n})\,\Gamma(d+1)\,\zeta(d+1)}{\beta^d} +
\frac{c^{\frac{d}{d+1}}}{1-c}\, \frac{y_0 M_{\rm
crit}(\vec{n})}{y(\vec{n})} \, . \Label{entropylargem}
\end{eqnarray}
At fixed $\vec{n}$, the limit of large $M$ is the limit of small
$c$ and consequently, from eq.~(\ref{cbeta}), large $\beta$. In
the latter case, we see that every term in (\ref{entropylargem})
goes to 0. This, of course, agrees with the intuition from the
microcanonical counting: when the number of summands grows too
large, the number of partitions of $\vec{n}$ quickly decreases,
reaching 0 when $M \propto |\vec{n}|$.

\subsection{Typical configurations of dual giant gravitons}
\llabel{sec:typicaldual}

As in Sec.~\ref{sec:entropy}, the results of the present section
afford an interpretation in terms of dual giant graviton
configurations. Indeed, the partition function
(\ref{partitiongen}) has been written down in \cite{sparks-Z}
specifically to count configurations of dual giant gravitons. In
particular, given the total momentum $\vec{P}$ on the maximal
torus of $X$, a typical distribution of momenta of the individual
dual giant gravitons is recovered using eq.~(\ref{meanocc2}) and
its generalization as in eq. ({\ref{generalize}). This is done, as
in Sec.~\ref{sec:entropy}, by selecting a special cycle
$\sigma_{\vec{\beta}} \equiv \sigma_{\vec{\nabla} f(\vec{P})} \in
\mathcal{C}$. One then sets the momentum component along
$\sigma_{\vec{\nabla} f(\vec{P})}$ for each dual giant graviton
using the (grand) canonical ensemble. The distribution of
3-momenta of the individual dual giant gravitons follows from this
procedure for essentially statistical reasons: the cycle
$\sigma_{\vec{\nabla} f(\vec{P})}$ is selected in such a way that
momentum along it splits into the three fundamental cycles of the
torus precisely in the ratio given by $\vec{P}$.

The different ensembles considered above have natural
interpretations in the language of dual giant gravitons. For
completeness, we dilate on this below:
\begin{itemize}
\item[1.] The analysis of Sec.~\ref{sec:mfree} applies to the case
where one sets the total momentum of the dual giant gravitons to
$\vec{P} \sim \vec{n}$ but leaves the total number of dual giant
gravitons unrestricted. The ensemble is then left to select its
own, optimal number of dual giant gravitons.
\item[2.] The analysis of Sec.~\ref{sec:mnon0} applies when one wishes
to fix the total momentum of the dual giant gravitons to $\vec{P}
\sim \vec{n}$ and the total number of dual giant gravitons to $M$.
The previous expressions are modified via eq.~(\ref{generalize}).
Several subcases are distinguished depending on the relation
between $M$ and $M_{\rm crit}(\vec{P})$, eq.~(\ref{mcrit}):
\begin{itemize}
\item[2.1] If $M \leq M_{\rm crit}(\vec{P})$ then eq.~(\ref{setalpha}) sets
the parameter $\alpha$ in a simple manner. In this case $M$ can be
thought of as fixing the \emph{exact} number of dual giant
gravitons, or the \emph{maximal} number of dual giant gravitons.
\item[2.2] If $M > M_{\rm crit}(\vec{P})$ fixes the \emph{maximal} number of
dual giant gravitons, the results of Sec.~\ref{sec:incl} apply. In
that case, the number of dual giants that maximizes entropy equals
$M_{\rm crit}(\vec{P})$.
\item[2.3] If $M > M_{\rm crit}(\vec{P})$ fixes the \emph{exact} number of
dual giant gravitons, the results of Sec.~\ref{sec:excl} apply.
\end{itemize}
\end{itemize}

\subsection{Summary: The universal features of heavy operators}

In all the theories we have considered, multi-trace operators and
their single-trace factors are charged under three $U(1)$s.   We
have shown that for a given total charge $\vec{n}$, the
distribution of charges of single-trace factors under a certain
$U(1)$ is universal.  This $U(1)$ is selected by a function
$f(\vec{n})$ on the dual toric cone which measures the minimal
volume in the cone cut out by a plane through $\vec{n}$.   Given a
triplet of charges $\vec{m}$, the charge of this particular $U(1)$
is computed by the inner product $t(\vec{m}) = (\vec{m} ,
\vec{\nabla} f(\vec{n}) )$.  Then the fraction of single trace
factors carrying U(1) charge $c_1 \, |\vec{n}|^{1/(d+1)} \leq
t(\vec{m}) \leq c_2 \, |\vec{n}|^{1/(d+1)}$ for any constants
$c_1$ and $c_2$ will be universal.    The distribution of charges
under the two residual $U(1)$s is not universal.   In addition the
total number of traces in a heavy multi-trace operator is
proportional to $t(\vec{n})^{d/(d+1)}$.  The proportionality
factor depends only on the dimension $d$ of the toric variety.

One concern about this result is that generic multi-trace
operators may not be orthogonal and hence we may wonder whether an
orthogonal basis enjoys the above universality properties.   To
examine this, we can imagine starting with one of the above
operators and then systematically producing an orthogonal basis by
a procedure like Gram-Schmidt.   By the arguments given above,
almost all the summands appearing in the orthogonal linear
combinations will have the same trace structure.   In this way, it
appears that the universality will extend to an orthogonal basis.
In any case, the dual giant graviton interpretation is free of
this caveat.   In that case, the universal properties include: (a)
the number of dual giants in a typical state with total momentum
$\vec{P}$, and (b) the fraction of the giants carrying a fixed
amount of momentum around a particular cycle denoted
$\sigma_{\vec{\nabla}f(\vec{P})}$.

\section{Dual black hole geometries}\setall
\llabel{sec:gravity}

In Sec.~\ref{sec:entropy} we have enumerated heavy scalar mesonic BPS operators with a given
charge vector $\vec{n}$. The structure of such operators exhibits many universal features,
which were described in Sec.~\ref{sec:structure-dist}, either in the trace basis or using the dual
giant graviton interpretation. The AdS/CFT correspondence anticipates that the dual
description of such states is in terms of black hole geometries
asymptotic to ${\rm AdS}_5$. We now turn to reinterpret our previous
results in terms of black hole physics.

\subsection{Dual black holes have vanishing horizons}

Here we argue that the dual black hole geometries have
semi-classically vanishing horizons. Since by the Bekenstein
relation the area is proportional to the entropy, $S = A/(4G_N)$, in
our case we have
\begin{equation}
S_{\textrm{grav.}} \sim \frac{\textrm{Vol}(X)}{l_p^5}\,
\frac{\rho_h^3}{l_p^3},
\end{equation}
where $\rho_h$ is the radius of the horizon and Vol$(X)$ is the
volume of the transverse Sasaki--Einstein manifold $X$. This
volume has been computed in \cite{sparks-volume}, and is
proportional to the volume of the 5-sphere as
\begin{equation}
\frac{\textrm{Vol}(X)}{\textrm{Vol}(S^5)} =
\frac{a_{\mathcal{N}=4}}{a_{\mathcal{N}=1}} \equiv h(a_i),
\end{equation}
where the constant of proportionality, $h(a_i)$, is the ratio of
the central charges of the two dual SCFTs. It was shown in
\cite{sparks-reeb} that the volume of $X$ is proportional to the
volume of the polytope $\Delta$ in the dual toric cone
$\mathcal{C}^*$ bounded by the plane $2(\vec{b},\,\vec{n}) = 1$:
\begin{equation}
\textrm{Vol}(X) = 6 (2\pi)^3 \,\textrm{Vol}(\Delta) = 2 (2\pi)^3
f((0,0,\frac{1}{6}))\, .
\end{equation}
The second equality follows from the discussion above
eq.~(\ref{reebexplicit}). As $f_{S^5}((0,0,\frac{1}{6})) =
\frac{1}{48}$, we have:
\begin{equation}
h(a_i) = 48 f((0,0,\frac{1}{6}))\, .
\end{equation}

Using the relation between string length and Planck length,
$l_p/l_s \sim g_s^{1/4}$, and the volume of the 5-sphere,
$\textrm{Vol}(S^5) \sim (g_sN)^{5/4} l_s^5$, we solve for the
horizon radius in string units as
\begin{equation}
\left(\frac{\rho_h}{l_s}\right)^3  \sim
\left(\frac{l_p}{l_s}\right)^3 \frac{l_s^5}{\textrm{Vol}(X)}
\left( \frac{l_p}{l_s} \right)^5 S_{\textrm{grav.}} \sim
\frac{g_s^2 S_{\textrm{grav.}}}{(g_sN)^{\frac{5}{4}} h(a_i)} \sim
\frac{ (g_sN)^{\frac{3}{4}}}{h(a_i)} \,
\frac{S_{\textrm{grav.}}}{N^2}. \Label{horizon}
\end{equation}
This indicates that the radius of the horizon is nonzero in string
units only if the entropy of the black hole scales as $N^2$ or
faster, where $N$ is the number of D-branes giving rise to the SCFT.

Naively, one would expect the dual geometries to correspond to
black hole microstates once the conformal dimension is of the
order $N^2$. This is because in other contexts black holes have
been described in terms of a collection of $N$ D-branes, each with
conformal dimension of the order $N$ \cite{superstar}. These
states do not have macroscopic horizons, however. To see this,
consider any lattice site $\vec{n}$ in the dual toric cone whose
conformal dimension is $\Delta$. The volume of any pyramid cut out
of the dual toric cone with a plane through $\vec{n}$ will grow as
$|\vec{n}|^d$. Therefore $f(\vec{n})\propto N^{2d}$ and by
eq.~(\ref{meinardus3}):
\begin{equation}
S(\vec{n})\propto N^{\frac{2d}{d+1}} = N^{2-\frac{2}{d+1}} \ll
N^2\,. \Label{sbound}
\end{equation}
We conclude that the entropy never grows sufficiently fast to
produce a horizon.

One might object that the above argument starts out treating $N$
as a large but finite parameter while the calculations leading to
eq.~(\ref{meinardus3}) assumed infinite $N$. The latter statement
is an immediate consequence of working with dual toric
\emph{cones}, which extend along infinite rays. By contrast, in
theories defined on finite stacks of D-branes, the ring of BPS
operators will be subject to some non-trivial relations, due to
the fact that for a rank $N$ matrix, the operator Tr ($X^{N+1}$)
can be written in terms of lower dimensional traces Tr($X$),
$\ldots$, Tr($X^{N}$). These relations are generated by Newton's
identities applied to minimal loop operators and begin to kick in
at distances of order $N$ away from the tip of the cone. As a
result, lattice points with $|\vec{n}| \gtrsim N$ in general fail
to produce algebraically independent trace factors. We conclude
that a finite $N$ may be thought of as defining a truncation of
the cone, so that all trace factors (partition summands) must fall
within a rough distance $N$ of the tip.

Happily,  below eq.~(\ref{meanocc3}) we argued that, in almost all
partitions, almost all summands satisfy $|\vec{m}| \propto
|\vec{n}|^{\frac{1}{d+1}} \propto N^{\frac{2}{d+1}}$, so that
$|\vec{m}| \ll N$ for $d>1$. Therefore, a generic partition does
not see the truncation by finite $N$. This establishes the
consistency of the argument leading to (\ref{sbound}). As an extra
observation, notice that the inequality (\ref{sbound}) is improved
as $d$ increases.

\paragraph{Finite horizons:}
Equation (\ref{horizon}) tells us how the entropy of a black hole solution should scale for the black hole to have a macroscopic
horizon. If we choose our conformal dimension to scale as $\Delta \sim N^{\gamma}$, we see that for a macroscopic horizon we need
\begin{equation}
N^2 \sim S \sim V^{\frac{1}{d+1}} \sim \Delta^{\frac{d}{d+1}} \sim
N^{\frac{\gamma d}{d+1}} \quad \Rightarrow \quad \gamma =
\frac{2(d+1)}{d} \to \frac{8}{3} \textrm{ for } d=3.
\end{equation}
Thus the states of interest to us have conformal dimensions $\Delta \sim N^{8/3}$.  Luckily, this is still not large enough to
invalidate the argument above: in almost all partitions almost all summands satisfy $|\vec{m}| \sim |\vec{n}|^{\frac{1}{d+1}}
\sim N^{\frac{\alpha}{d+1}} \sim N^{2/3}$, so that $|\vec{m}| \ll N$. It is interesting to note  that $\gamma$ decreases as $d$
increases in keeping with the fact that less supersymmetry makes it  easier to develop a horizon.\footnote{~It is unclear to us
whether such heavy operators will have a classical spacetime description in terms of an asymptotically $AdS_5\times S^5$ metric,
or whether this asymptotics will be modified.}

\paragraph{Gravity duals of heavy operators:} For states carrying non-vanishing R-charge in the superconformal field theory, the dual metrics were given in \cite{kim}, who used the consistent truncation discovered in \cite{buchel}. This R-charge is realized in supergravity by turning on angular
momentum along the cycle defined by the Reeb vector. Denoting by $\partial_\phi$ the Killing vector field along the Reeb vector, the metric reads
\begin{equation}
ds^2 = -\frac{1}{4} H^{-2} F dt^2 + H [ F^{-1} dr^2 + r^2
ds_{S^3}^2] + ds_{Y}^2 + (d\phi + A)^2,
\end{equation}
where $Y$ is a four dimensional K\"ahler--Einstein manifold, and
\begin{eqnarray}
H & = & 1+\frac{q}{r^2}, \nonumber \\
F & = & 1+r^2H^3, \\
A & = & \frac{1}{2}H^{-1}dt\,. \nonumber
\end{eqnarray}
The gauge field $A$ is responsible for the non-vanishing R-charge.
These metrics have manifestly vanishing horizons, in agreement
with our field theory counting and stretched horizon
considerations. They carry only R-charge and are not charged under
any extra $\U(1)$s. This is equivalent to choosing our charge
vector $\vec{n}$ to point in the $z$-direction (in the standard
presentation), as dictated by (\ref{maxR}).

For $\mathcal{M}=\IC^3$, (singular) supersymmetric configurations
with vanishing horizons have been known to exist for any value of
the three $\U(1)$ R-charges, as long as the system remains BPS
\cite{sabra}. This statement matches our observations for any
charge vector $\vec{n}$.

Since our methods and results apply to general toric varieties
$\mathcal{M}$ with arbitrary $\U(1)$ charges turned on, any
comparison with gravity would require the addition of these extra
quantum numbers (nonzero flavor charges). To our knowledge such
solutions have not been written down in the literature. We claim
that the relevant geometries, when discovered, will be
characterized by vanishing horizons.

\section{Discussion}\setall
\llabel{discussion} In this paper we developed a geometric way of
counting scalar mesonic operators in $\mathcal{N}=1$ toric quiver
gauge theories based on the one-to-one correspondence between
(equivalence classes of) gauge invariant operators and lattice
points in the corresponding dual toric cone. This correspondence also
allowed the recent development of the so-called plethystic program
for counting the same quantities \cite{plethystic}.

Our geometric considerations allowed us to analyze the structure
of mesonic operators in the large charge limit using statistical
arguments. This was accomplished by working with the appropriate
partition functions (ensembles), and realizing that the set of
states having the same occupation numbers form planar surfaces in
the dual toric cone, inducing a natural grading on it. This
connection allowed us to find the direction in the dual toric cone that had the most rapid growth of entropy.

In the large charge limit, the grading becomes a continuous
variable $y$.  We were able to use this  to encode the structure of gauge
invariant operators in terms of curves $x_d(y)$, whose properties
only depend on the dimension $d$ of the initial toric
variety.\footnote{~In other words, the details about the precise
nature of the toric variety are suppressed by negative powers of
the charge, or equivalently, are suppressed in $N$.} Here, $x_d(y)$
stands for the ratio of traces in a given multi-trace operators
having grading value $\tilde{y} < y$. Similar mathematical
functions control the distribution of all $U(1)$ charges among the
traces of the heavy multi-trace operator.

Our statistical analysis shows that there exists a universal
structure, shared by almost all heavy operators. In this sense,
given an arbitrary constraint like fixed R-charge, there exists a
typical state close to which almost all other states lie; these
can be thought of as master states in the field theory. This
provides a generalization, for any $\mathcal{N}=1$ toric quiver
gauge theory, of the results and philosophy reported in
\cite{babel} for the half-BPS sector of $\mathcal{N}=4$ SYM.

We mainly worked on the field theory side, and according to the
AdS/CFT duality appropriate combinations of these operators are
expected to be dual to microstate geometries on the gravity side.
Unfortunately, the exact form of these geometries is not known for
a general quiver.\footnote{~Recently, for the special case where
the corresponding toric variety is $\mathbb{C}^3$, the equations
describing the moduli space of such candidate microstate
geometries were constructed in \cite{microstates}.} Should these
geometries be discovered in the future, our methods could be used
to analyze the emergence of semi-classical gravity from the
fundamental microstates in the spirit of
\cite{us-entropy}.\footnote{~Some recent work in the literature
concerns the emergence of locality and spacetime from a study of
the strong coupling dynamics of the gauge theory dual to the
conifold \cite{berenstein2007}. It would be of interest to apply
these techniques both to our heavy operators and to more general
$\mathcal{N}=1$ setups.}

Even if we would know the explicit candidate microstate
geometries, we know that their coarse-grained description would be
interpreted in terms of distributions of giant gravitons or dual
giants in AdS$_5 \times X$. All our results can actually be
reformulated in terms of dual giants due to the isomorphism of
Hilbert spaces discussed in Section
\ref{sec:dualphases}.\footnote{~As discussed in appendix
\ref{ap:amoeba}, the existence of the identity between partition
functions \eqref{fund-pleth} suggests that a similar
interpretation in terms of giants should also be possible, and
would be more naturally formulated in terms of fermions.} Since
our universality conclusions also apply in the description in
terms of dual giant gravitons, this strongly suggests that the physical
picture of black hole microstates  advocated in \cite{babel} will also hold for all the
systems described in this paper.

We have only worked at a classical level, and giants and dual
giants are quantum mechanical objects. Thus, to establish the
validity of our analysis, we need to show that the wavefunction of
the configuration of D-branes is strongly localized, indicating
that the system has a classical description. We established this
localization for the case of giant gravitons on $S^5$ in appendix
\ref{typicalgiant} and expect it to hold in a general setting.

In Section~\ref{sec:gravity}
we showed that to produce a macroscopic horizon,  we would have to consider
BPS mesonic states
with conformal dimensions of at least of order $N^{8/3}$,
for the case of a three dimensional toric variety, i.e., for
string theory on $\ads{5} \times X$.  It would be interesting to
understand what these states look like on the gravity side, and
why such large conformal dimensions are needed to create horizons.
Naively one would expect a black hole to arise from a collection
of $\mathcal{O}(N)$ D-branes and thus have a conformal dimension
$N\cdot N \sim N^2 \ll N^{8/3}$. However, whatever the
gravitational description, the structure of these states can be
studied using our methods, barring a breakdown of the canonical
ensemble itself.\footnote{~The ensemble might conceivably break
down due to a Hagedorn-like growth in the number of states; it
might be interesting to study this possibility further.}

We would also like to point out that the case of a four (complex)
dimensional toric variety might be interesting to study. In this
case, the conformal dimension required to produce a macroscopic
horizon is smaller, $N^{5/2}$, and the field theory is simpler.
This is because increasing the (real) dimension of the toric
variety by two, correspondingly decreases the dimension of the
transverse Minkowski space by two, and after the near horizon
limit this yields string theory on $\ads{3} \times \tilde{X}$,
where $\tilde{X}$ is is the seven-dimensional base of the toric
variety. This is naturally expected to be dual to a
$(1+1)$-dimensional CFT.

Finally, we argued that even though the trace basis is not
orthogonal for large values of the $\U(1)$ charges, the typicality
we found would also be present in an orthogonal basis. However, it
would be interesting to work explicitly with such an orthogonal
basis, analogous to the basis of Schur polynomials for the
half-BPS sector of $\mathcal{N}=4$ super-Yang--Mills \cite{cjr}.
Such a basis was recently provided in \cite{ramgoolam} for the
chiral ring of $\mathcal{N}=4$ super-Yang--Mills. It would be
interesting to extend that analysis and techniques to chiral rings
of arbitrary $\mathcal{N}=1$ superconformal field theories.

We expect that last point to be relevant for a comparison with
gravity. This is because a generic heavy multi-trace chiral
operator has no natural giant graviton interpretation
\cite{us-entropy} and we already know that Schur polynomials
played a crucial role in matching the analysis of typicality in
gauge theory with gravity \cite{babel}.

\paragraph{Acknowledgements:}
The authors wish to thank Bo Feng, Kris Kennaway, Dario Martelli,
Yuji Tachikawa and Brian Wecht for interesting conversations.
VB, BC and KL were supported in part by the DOE under grant
DE-FG02-95ER40893, by fellowship from the Academy of Finland (KL),
and a Dissertation Completion Fellowship from the University of
Pennsylvania (BC).  VB is also grateful for support as the Helen
and Martin Chooljian member of the Institute for Advanced Study.
YHH is indebted to the FitzJames Fellowship of Merton College, Oxford.

\appendix
\section{Exact multiplicities of operators with irrational R-charges}
\setall
\llabel{irrational}

In Section~\ref{partplanar} we showed that the combined entropy of
operators with co-planar charge vectors $\vec{n}$ is a function
(eq.~(\ref{meinardus2})) of the volume which the plane cuts out of
the dual toric cone. The simplest application of this statement is
to evaluate the total entropy of all operators of fixed R-charge
$R_0$. The charge vectors satisfying this constraint live on the
plane
\begin{equation}
(\vec{b},\,\vec{n}) = \frac{3}{2}R_0\, .\Label{constrainR}
\end{equation}
The volume of $\mathcal{C}^*$ bounded by the plane determines the
total entropy via eq.~(\ref{meinardus2}).

The above statement is valid in the following two cases:
\begin{itemize}
\item When the Reeb vector $\vec{b}$ contains only one $\mathbb{Z}$-linearly independent component,
\item When one wishes to impose (\ref{constrainR}) up to some small
finite accuracy $\epsilon$.
\end{itemize}
Suppose, however, that among the components of the Reeb vector
there are $k>1$ $\mathbb{Z}$-linearly independent ones. Denote
them by $r_i,\,i=1,\ldots,k$. One example of that is $dP_1$, where
$\vec{b}=(0,\,4-\sqrt{13},\,3)$ and $4-\sqrt{13},\,3$ are
$\mathbb{Z}$-linearly independent. Suppose further that one is
interested in constraining the R-charge at \emph{exactly} $R_0$.
We may decompose $R_0$ and $\vec{b}$ into summands proportional to
$r_i$:
\begin{eqnarray}
\vec{b} & = &
r_1\,(b_1^1,\,b_2^1,\,b_3^1)+\ldots+r_k\,(b_1^k,\,b_2^k,\,b_3^k)\nonumber\\
R_0     & = & r_1\,R_0^1 +\ldots + r_k\,R_0^k\, .
\end{eqnarray}
If no decomposition of $R_0$ is possible, then no state with
$R=R_0$ exists and computing the entropy is meaningless.
Otherwise, the constraint (\ref{constrainR}) decomposes into $k$
independent ones:
\begin{equation}
(\vec{b}^i,\,\vec{n}) = \frac{3}{2}R_0^i\quad\quad i=1,\ldots,k\,
.\Label{constrainRk}
\end{equation}
Therefore the locus of charge vectors \emph{exactly} matching
$R_0$ is given by the region of intersection of $\mathcal{C}^*$
and $k$ planes (\ref{constrainRk}). In other words, it is the
section of a certain $(d-k)$-hyperplane contained in the dual
toric cone.

For $d=3$, three cases are distinguished:
\begin{enumerate}
\item $k=1$. The locus $R=R_0$ is a plane. The combined entropy
is determined by the volume of $\mathcal{C}^*$ contained under the
plane as explained in Section~\ref{partplanar}.
\item $k=3$. The locus $R=R_0$ is a point. The entropy is
determined by the minimal volume of $\mathcal{C}^*$ that may be
cut out with one plane through the point $R=R_0$. This was
explained in Section~\ref{partind}.
\item $k=2$. The locus $R=R_0$ is a line. This case, along with all
others with $1<k<d$, is treated in the next subsection.
\end{enumerate}

\subsection{Multiplicities of collinear vectors}

The argument of Section~\ref{partind} is readily adapted to
$1<k<d$. The locus of points of interest is a $(d-k)$-hyperplane
satisfying $k$ equations (\ref{constrainRk}). Consider the family
of $(d-1)$-hyperplanes $H_{\vec{s}}$ containing the locus $R=R_0$.
Here $\vec{s}$ is the normal to the hyperplane, which is specified
by $k-1$ free parameters. In particular, $\vec{s}$ is a linear
combination of the vectors $\vec{b}^i$ whose normalization is
immaterial and may be scaled to 1.

For each such $\vec{s}$, consider the family of $(d-1)$-planes
parallel to it and proceed as in Section~\ref{partplanar}. The
total entropy of the points $\{R=R_0\}$ is bounded from above by
\begin{equation}
S(R_0)\, \leq \min_{\vec{s}=\sum_{i=1}^k a_i\,\vec{b}^i}
S(p_{\vec{s}})\,.
\end{equation}
Here $S(p_{\vec{s}})$ denotes the total multiplicity of all charge
vectors contained in $p_{\vec{s}}$, which may be obtained from
eq.~(\ref{meinardus2}). Assume that every $(d-k)$-hyperplane
$\{R=R'_0\}$ in $p_{\vec{s}_{\rm min}(R_0)}$ has its own minimal
$\vec{s}_{\rm min}(R'_0)$. Since the region $\{R=R'_0\}$ is
contained in $p_{\vec{s}_{\rm min}(R_0)}$, its minimum must be
strictly smaller than the entropy of the $(d-1)$-plane
$p_{\vec{s}_{\rm min}(R_0)}$. This applies to every locus
$\{R=R'_0\neq R_0\}$, which implies that:
\begin{equation}
S(R_0)\, \approx \min_{\vec{s}=\sum_{i=1}^k a_i\,\vec{b}^i}
S(p_{\vec{s}})\,. \Label{deffk}
\end{equation}
This equation applies to all $1\leq k\leq d$ and naturally
generalizes eq.~(\ref{define-f}).

As $k$ grows, one is able to derive more information from a single
exact measurement of the R-charge. Such successive specifications
narrow down the microcanonical ensemble of states $\{R=R_0\}$,
which should result in a decrease of the multiplicities computed
in eq.~(\ref{deffk}). We may see that this is indeed the case from
the schematic relationship
\begin{eqnarray}
S(R_0)|_{k=d\phantom{-1}} & \approx &
\min_{\vec{s}=\sum_{i=1}^d a_i\,\vec{b}^i} S(p_{\vec{s}}) \nonumber \\
\leq\, S(R_0)|_{k=d-1} & \approx &
\min_{\vec{s}=\sum_{i=1}^{d-1} a_i\,\vec{b}^i} S(p_{\vec{s}})
\,\,\leq\, \ldots \nonumber \\
\leq\, S(R_0)|_{k=1\phantom{-d}} & = &
\phantom{\min_{\sum_{i=1}^{d-1} \vec{s}=a_i\,\vec{b}^i}}
S(p_{\vec{b}})\,. \Label{hierarchy}
\end{eqnarray}
The multiplicities obtained from specifying the charge vector
$\vec{n}$ with a growing number of constraints differ only in the
domain of minimization in eq.~(\ref{deffk}). In particular, they
are of the same order of magnitude.

\section{Mapping to the typical giant configuration}
\setall \Label{typicalgiant} In Section~\ref{sec:typicaldual} we
interpreted our analysis as giving a typical configuration of dual
giant gravitons. However, all these configurations are at heart
quantum mechanical, and to analyze typicality we should verify
that the wavefunction of the typical configuration is sharply
peaked, thereby allowing one to associate it to a semiclassical
collection of D-branes.  In this appendix we will argue that this
indeed is the case: the typical states constructed above can be
associated to a typical configuration of giant gravitons on the
gravity side. For simplicity, we will deal with giant gravitons on
$S^5$; we expect the conclusions to be valid in a more general
setting as well.

It was shown by Mikhailov \cite{mikhailov} and Beasley
\cite{beasley} that these brane configurations can be given in
terms of zero-loci of polynomials of the form
\begin{equation}
P(z_1,z_2,z_3) = \sum_{\vec{n}} c_{n_1n_2n_3} z_1^{n_1} z_2^{n_2} z_3^{n_3},
\end{equation}
where $\vec{n}$ runs over $\mathbb{N}^3$. The worldvolumes of the
branes are given by the intersection of the surface $P(\vec{z}) =
0$ with the unit $S^5$ in $\mathbb{C}^3$, where the $S^5$ is
identified with the one in $\textrm{AdS}_5 \times S^5$. The phase
space of brane configurations is thus spanned by $\{ c_{n_1n_2n_3}
\}$ and is topologically $\mathbb{CP}^{\infty}$.

This space was quantized in \cite{minwalla} using geometric
quantization. First one defined a new set of coordinates that
eliminated the pathological regions in coordinates $\{
c_{n_1n_2n_3} \}$, i.e., eliminated polynomials that don't
intersect the unit $S^5$ and identified polynomials that have the
same intersection with the $S^5$. The new coordinate system is
given as
\begin{equation}
w_{n_1n_2n_3} = f_{|n_1+n_2+n_3|}(\vec{c}) c_{n_1n_2n_3}, \Label{ctow}
\end{equation}
where $f_{|n_1+n_2+n_3|}$ is a smooth function. Upon quantization
the Hilbert space of brane configurations was found to be the
space of degree $N$ holomorphic polynomials in $w_{n_1n_2n_3}$,
which is isomorphic to the Hilbert space of $N$ free bosons in a
three dimensional harmonic oscillator, and therefore has the
partition function
\begin{equation}
Z = \prod_{j=1}^{N_c} \frac{1}{\left( 1-e^{-\beta j} \right)^{\frac{1}{2}j(j+1)}}. \Label{Z3D}
\end{equation}
Above, we've restricted the highest excitation any particle can
have to be given by the constant $N_c$; physically this corresponds to fixing the the number of giant gravitons \cite{babel}.
This is because the highest excitation gives the degree of the polynomial $P(z_1,z_2,z_3)$ and
therefore the maximal number of D-branes. In the following we'll
also work in the high temperature regime, $\beta\to 0$.

We can now write down the wave function of the typical state. Since the energy levels factorize, we have
\begin{equation}
\Psi^{\textrm{typical}}(\vec{w}) = \prod_{j=0}^{N_c} \psi_j (\vec{w}), \Label{typical1}
\end{equation}
where $\psi_j$ is the contribution from particles with energy $j$
to the total wavefunction. Since the wave function has to be
symmetric both under the exchange of particles and under the
exchange of dimensions, we can write
\begin{equation}
\psi_j(\vec{w})  =  \left[ \sum_{\substack{n_x,n_y,n_z = 0 \\ n_x+n_y+n_z = j}}^{j} w_{n_x,n_y,n_z} \right]^{k_j}.
\Label{typical2}
\end{equation}

Since the moduli space is $\mathbb{C} \mathbb{P}^{\infty}$, the
coordinates have to satisfy $\sum_{n_x,n_y,n_z}
|w_{n_x,n_y,n_z}|^2 = 1$. If we define the contribution from each
energy level $j$ as $R_j^2$, we can write this condition as
\begin{equation}
\sum_{\substack{n_x,n_y,n_z = 0 \\ n_x+n_y+n_z = j}}^{j} |w_{n_x,n_y,n_z}|^2 = R_j^2, \quad \textrm{with } \sum_{j=0}^{ N_c}
R_j^2 = 1.
\end{equation}
For any given $R_j$, we know that the maximum of
$|\sum_{\substack{n_x,n_y,n_z = 0 \\ n_x+n_y+n_z = j}}^{j}
w_{n_x,n_y,n_z} |$ occurs when all the $w$'s are equal, and since
the number of states at energy level $j$ is $\binom{j+2}{2}$, we
see that the correct value is
\begin{equation}
w_{n_1n_2n_3} \equiv w_j = \sqrt{\frac{R_j^2}{d_j}}, \quad \forall \, \, n_1 + n_2 + n_3 = j.
\end{equation}
(Up to an overall phase that is not relevant to us.) Thus at the maximum the wavefunction  can be written as
\begin{equation}
\Psi(\vec{w}) = \prod_{j=0}^{N_c } \left[ \binom{j+2}{2}\, w_j \right]^{k_j} = C \prod_{j=0}^{N_c} R_j^{k_j},
\end{equation}
where the normalization constant $C$ is not important and will be
dropped. We wish to maximize $|\Psi|^2$ with the constraint $\sum
R_j^2 =1$, and to do this we eliminate $R_0$ by solving the
constraint, yielding
\begin{equation}
|\Psi(\vec{w})|^2 = \left( 1 - \sum_{k=1}^{N_c} R_k^2 \right)^{k_0} \prod_{l=1}^{N_c} R_l^{2k_l}.
\end{equation}
After some algebra one finds the maximum of this at
\begin{equation}
R_j = \sqrt{\frac{k_j}{N}}, \quad \forall \, \, j,
\end{equation}
which by construction satisfies $\sum R_j^2 = 1$. This is the only
local extremum, and since the wavefunction vanishes on the `edges'
where $w_j = 0$ for some $j$, this is also the global maximum.

Finally, we compute the second derivative of the wavefunction at
the maximum. After some algebra this yields
\begin{equation}
\left. \frac{1}{|\Psi(\vec{w})|^2} \frac{\partial^2 |\Psi(\vec{w})|^2}{\partial R_j \partial R_i}\right|_{max} = - 4N
(\frac{\sqrt{k_i k_j}}{k_0} + \delta_{ij}).
\end{equation}
Since in the large temperature limit $\beta \to 0$ the free energy
is minimized when the entropy is maximized, we see that for the
typical state the particles are evenly distributed among all
possible states\footnote{~This is also easy to derive from
(\ref{Z3D}) more rigorously.} , and therefore the number of
particles on energy level $j$ is  $k_j \propto \binom{j+2}{2}$.
Thus we see that in the limit $N \to \infty$ the wavefunction is
sharply peaked around the maximum, and  we can associate the
typical state with the location of the maximum in the phase space,
$\vec{w}^{max}$. The map (\ref{ctow}) then guarantees that there
exists a corresponding point $\vec{c}^{max}$ specifying the
typical configuration of giant gravitons. We expect this
configuration to correspond to the superstar \cite{superstar}, but
have been unable to verify this as the explicit form of the
funtion $f_{|n_1+n_2+n_3|}(\vec{c})$ in (\ref{ctow}) is not known.

\section{Limiting Curves and Amoebae?}
\setall
\label{ap:amoeba}

\newcommand{\eref}[1]{(\ref{#1})}

There have recently been developments in topological strings using the statistical physics of
so-called melting crystals \cite{melting,melting2}. Analyses of typical states in black
hole physics have involved the use of similar statistical
mechanics techniques. In this brief appendix we would like to
comment on the similarities and differences between the two and
make some speculative remarks.

In the melting crystal studies one considers geometric
quantization of a \emph{single particle} in $\IC^m$. Its spectrum
is that of a harmonic oscillator in $m$ dimensions, and as such,
the associated partition function is related to a classical
partition function. For example, for $\IC^3$, there is a
one-to-one map between quantum harmonic oscillator states and 3-D
Young tableaux. If we associate each tableaux with a distribution
of boxes in an octant of $\IR^3$, the statistical mechanics
provides the limiting shape for the boundary of the crystal as
a function of the string coupling, which plays the role of the
temperature in this set-up.

This discussion is conceptually very reminiscent of the derivation of the typical state in the half-BPS sector of $\cN=4$ SYM. There is a one-to-one correspondence between {\it $N$-particle} fermionic quantum states in a 1-dimensional harmonic oscillator and 2-dimensional Young tableaux. Statistical mechanics
and a large $N$ limit allow the derivation of a limit shape $y(x)$ which describes the average excitation $(y)$ of the particle $x$.

Though the techniques are the same, the physics looks rather
different. The plethystic methods and the geometric quantization
of the classical moduli space of dual giants provide a clue as to
where a relation may exist. Let us consider the isomorphism,
which is a cornerstone of the plethystic program:
\begin{equation}
  g_N(t,\,\eM) = g_1(t,\,\eM^N/S_N)\,.
 \label{fund1}
\end{equation}
This states that the number of mesonic scalar chiral primary gauge
invariant multi-trace operators constructed out of building blocks
that define $\eM$ is equal to the number of single trace operators
in a different manifold, the symmetric product $\eM^N/S_N$. In
terms of dual giant gravitons, this has a very natural
interpretation: the Hilbert space of $N$ dual giants is equivalent
to the symmetric space of the Hilbert space spanned by a single
giant, which is the set of holomorphic functions in $\eM$.
Therefore, \eqref{fund1} provides a bridge between partition
functions of \emph{$N$ particles} and partition functions of a
\emph{single particle} in a different space.

Consider the particular case of $\eM=\IC$. The Hilbert space of a
single particle is that of a harmonic oscillator in one dimension.
When we consider the Hilbert space of a single particle in
$\IC^N/S_N$, all quantum numbers in each of the copies appearing
in the symmetric product have to be different: this is effectively
implementing the Pauli exclusion principle, and so we conclude
that we are equivalently dealing with $N$ {\it fermions} in a one
dimensional harmonic oscillator. The limiting curve, exhibited in
eq.~(44) of \cite{babel}, is given by
\begin{equation}\label{C-limitcurve2}
a e^{\beta x} + b e^{-\beta y} = 1 \ , \qquad a = e^{-\beta N},
\quad b = 1- a \
\end{equation}
and reproduces the curve $x_1(y)$ of eq.~(\ref{defx}) up to a
linear re-definition of coordinates. Inspecting Section~4.1 of
\cite{melting2} we see the emergence of \eref{C-limitcurve2}
therein. Equivalently, in the language of \cite{dimer},
\eref{C-limitcurve2} is the boundary of the amoeba of $\IC^3$,
which, in turn, is equivalent to the limiting shape of
constructing three-dimensional Young tableaux.

We can now understand the space where the limiting curve $y(x)$
lives. Indeed, $x$ can be understood as parameterizing the copy
$x$ in the symmetric product $\IC^N/S_N$ (or the particle $x$ in
the dual giant language) whereas $y$ is just describing the
excitation of that copy (particle). In the semiclassical limit $x$
is a continuum variable, and we obtain a curve on a two
dimensional plane. This curve describes a single particle living
not in $\IC^2$, but rather in the symmetric product $\IC^N/S_N$
with $N\to \infty$.

To recapitulate, we have that (1) the curve of typicality of
half-BPS states of $\cN=4$ SYM identifies with (2) the limiting
curve of the topological A-model on $\IC^3$. We point out that the
underlying geometries of the two differ: the fundamental
generating function (i.e., for the single-trace operator) of (1),
in the language of the plethystic program, is the Hilbert series
for $\IC$ while the curve in (2) is the amoeba for $\IC^3$.
Nevertheless, the emergence of the same curve in two different
counting problems suggests that there may be a deeper connection,
possibly persisting to other geometries, and hence other $\CN=1$
theories, which we should explore.


\end{document}